\pdfoutput=1
\documentclass[
reprint,oneside, twocolumn,
nofootinbib,
amsmath,amssymb,
aps,
]{revtex4-2}

\usepackage{graphicx}
\usepackage{dcolumn}
\usepackage{hyperref}
\usepackage{slashed}
\usepackage{cancel}
\usepackage{upgreek}
\usepackage{wrapfig}
\usepackage{bbm}
\usepackage[svgnames,dvipsnames,x11names]{xcolor}
\usepackage{bm}
\usepackage{float}
\usepackage{geometry}
\usepackage{yfonts}
\usepackage{sidecap}
\usepackage{anyfontsize}
\usepackage{dsfont}
\usepackage{tikz}
\usepackage{braket}
\usepackage{mathtools}
\usepackage{tensor}
\usepackage{soul}

\hypersetup{colorlinks=true,
	breaklinks=true,
	pdfstartview=Fit,
	linkcolor=purple,
	citecolor=purple,
	urlcolor=purple}

\graphicspath{{Fig/}}

\newgeometry{top=2.5cm, bottom=2.5cm, left=2cm, right=2cm}

\def\Mpl{M_{\text{Pl}}} 
\def\calT{\mathcal{T}} 
\def\d{\mathrm{d}}
\def\kp{k_{\text{p}}}
\def\g{\gamma}
\def\D{\mathcal{D}}

\def\Im{\mathrm{Im}\,}
\def\Res{\mathrm{Res}\,}

\definecolor{pyblue}{RGB}{31, 119, 180}
\definecolor{pyorange}{RGB}{255, 127, 14}
\definecolor{pygreen}{RGB}{44, 160, 44}
\definecolor{pymiddle}{RGB}{105, 136, 79}
\definecolor{pyred}{RGB}{214, 39, 40}

\begin{document}

\preprint{APS/123-QED}

\title{Parity Violation from Emergent Non-Locality During Inflation
}

\author{Sadra Jazayeri,$^{1}$ Sébastien Renaux-Petel,$^{1}$ Xi Tong,$^{2, 3}$ Denis Werth$^{1}$ and Yuhang Zhu$^{2, 3}$}

\affiliation{$^{1}$ Institut d'Astrophysique de Paris, UMR 7095 du CNRS et de Sorbonne Universit\'e, 98 bis bd Arago, 75014 Paris, France}

\affiliation{$^{2}$Department of Physics, The Hong Kong University of Science and Technology, Clear Water Bay, Kowloon, Hong Kong, P.R. China}

\affiliation{$^{3}$The HKUST Jockey Club Institute for Advanced Study, The Hong Kong University of Science and Technology, Clear Water Bay, Kowloon, Hong Kong, P.R. China}

\date{\today}

\begin{abstract}
	
Parity violation in the early universe holds great promise for uncovering new physics. In particular, the primordial scalar four-point correlation function is allowed to develop
a parity-violating component when 
massive spinning particles coupled to a helical chemical potential are present during inflation. In this paper, we explore the rich physics of such a parity-violating trispectrum in the presence of a reduced speed of sound for the Goldstone boson of broken time translations. We show that this signal can be significantly large while remaining under perturbative control, offering promising observational prospects for future cosmological surveys. In the limit of a reduced sound speed, the dynamics admits an effective \textit{non-local} description organized as a time-derivative expansion. This reveals that parity violation arises due to emergent non-locality in the single-field effective theory. At leading order, this effective theory yields a compact trispectrum template, written in terms of elementary functions. We then conduct a comprehensive analysis of the kinematic dependence of this parity-violating trispectrum and reveal new features. In addition to the low-speed collider resonance, we find a \textit{new class of signals} lying in the internal soft-limit of the correlator. This signal is characterized by an oscillatory pattern periodic in the momentum ratio, with a frequency determined by the speed of sound and the chemical potential, making it drastically distinct from the conventional cosmological collider signal.

\end{abstract}

\maketitle


\section{Introduction}

At the macroscopic scale of our everyday experiences, the mirrored world appears to be remarkably indistinguishable from our own. Therefore, parity violation---namely spatial coordinates inversion---in the weak interactions at microscopic scales came as a total surprise in the 1950s \cite{Lee:1956qn,Wu:1957my}. After the subsequent establishment of the Standard Model (SM) \cite{Glashow:1961tr,Weinberg:1967tq,Salam:1968rm}, we now become acquainted with the notion of parity asymmetry in particle physics. It is thus natural to continue asking: is the world parity-invariant on cosmological scales? Recent observations of the Cosmic Microwave Background (CMB) birefringence \cite{Minami:2020odp,Diego-Palazuelos:2022dsq} and the Large Scale Structure (LSS) of the universe \cite{Hou:2022wfj,Philcox:2022hkh} seem to suggest that parity violation is indeed possible on cosmological scales.

\vskip 4pt
The question naturally arises as to which observables are sensitive to violation of parity. 
In the tensor sector, parity violation can be revealed already at the level of the two-point correlation function, as each helicity of the graviton may exhibit distinct power. By contrast, for scalar observables, the lowest-order connected correlation function sensitive to parity violation is the four-point function, known as the primordial trispectrum, on which we focus in this work.
The reason for this is simple. The kinematic configurations of lower-point correlation functions are \textit{planar}, and are therefore blind to any information about parity since their mirrored images are always related to themselves via a spatial rotation. One needs at least four non-planar points to establish chirality. In Fourier space, because a parity transformation on Hermitian operators is equivalent to taking the Hermitian conjugate $\mathcal{O}_{\bm{k}}^\dagger = \mathcal{O}_{-\bm{k}}$, parity violation in the trispectrum manifests itself as a parity-odd imaginary component. Although it remains unclear whether the recent hint of parity violation in the sky is of primordial origin, see e.g.~\cite{Cabass:2022oap, Philcox:2023ffy,Philcox:2023ypl}, it is natural to expect at least some level of parity violation in the primordial universe. This is because the observed cosmological structures, such as density fluctuations in the CMB and LSS, are believed to be seeded by \textit{microscopic} quantum fluctuations during inflation, where parity is not necessarily exact.

\vskip 4pt
Parity violation is forbidden in tree-level cosmological correlators for vanilla single-field inflation \cite{Liu:2019fag, Cabass:2022rhr}. Physically, this can be understood as a consequence of unitarity and scale invariance, while mathematically it directly follows from the simple analytic structure of the correlator time integrals. However, there are several ways to circumvent this no-go theorem. These include, for example, breaking scale invariance \cite{Cabass:2022rhr}, considering non-linear dispersion relations \cite{Cabass:2022rhr, Creque-Sarbinowski:2023wmb}, or going to loop level \cite{Lee:2023jby}. More naturally, parity violation can also be generated by additional massive particles in the early universe. Such particles are generically present during inflation as they can come from supersymmetry breaking \cite{Baumann:2011nk, Delacretaz:2016nhw,Alexander:2019vtb}, SM uplifting \cite{Chen:2016uwp, Kumar:2017ecc, Hook:2019zxa, Chen:2016nrs, Chen:2016hrz}, isocurvature sector \cite{Chen:2009zp, McAllister:2012am, Lu:2021gso, Chen:2023txq}, or arise as Kaluza-Klein modes or higher-spin stringy states \cite{Rindani:1985pi, Aragone:1987dtt, Kumar:2018jxz, Baumann:2014nda}. As a consequence, parity violation appears as a compelling opportunity to discover new physics beyond the minimal and unavoidable scalar and tensor fluctuations of the metric. One natural possibility of inducing parity violation with massive spinning bosonic particles is by their coupling to a \textit{helical chemical potential}, see e.g.~\cite{Adshead:2015kza, Wang:2019gbi, Wang:2020ioa, Sou:2021juh, Wang:2021qez, Tong:2022cdz, Qin:2022lva, Qin:2022fbv}. However, the amplitude of the corresponding parity-odd trispectrum from exchanging such particles of mass $m$ is typically small for heavy fields. This is because the no-go theorem states the whole tower of parity-odd effective field theory (EFT) operators in the $\Box/m^2$-expansion are screened at tree-level by scale invariance. Thus the corresponding parity violation signals are either loop suppressed, slow-roll suppressed, or exponentially small, i.e.~$\mathcal{O}(e^{-\pi m/H})$, with $H$ denoting the Hubble scale during inflation. Therefore, this parity-odd component becomes undetectable for large enough masses.

\vskip 4pt
In this paper, we explore a natural framework consisting of an additional spin-1 massive field with a chemical potential present during inflation, and consider the theoretically motivated scenario in which the Goldstone boson of broken time translations propagates with a reduced speed of sound~$c_s~\ll~1$. We show that a large parity-violating primordial trispectrum can be produced while remaining under perturbative control, and that its amplitude is enhanced to $\mathcal{O}(e^{-c_s m/H})$. In the small sound-speed limit~$c_s~\ll~(m/H)^{-1}$, the signal is not suppressed and the dynamics admits a single-field effective description that is \textit{non-local in space}, in which the heavy field is integrated out in a non-standard way. This regime is known as the low-speed collider regime, see \cite{Jazayeri:2022kjy,Jazayeri:2023xcj} for more details. We show that the resulting parity-odd component of the trispectrum is completely fixed by the singularity of the non-local interactions, bypassing the no-go theorem even in the single-field limit. The derived parity-odd template consists of simple rational and exponential functions, thereby greatly reducing the computational costs of future non-Gaussianity data analysis. We then conduct a thorough analysis of the kinematic dependence of the identified signal. In particular, we show that the internal mildly-soft 
kinematic configurations exhibit
a resonance peak that is identified as the low-speed collider signal \cite{Jazayeri:2022kjy,Jazayeri:2023xcj}. Its location is fixed by a combination of the sound speed, the mass and the chemical potential. In addition, we find a new family of oscillatory signals that are periodic in the momentum ratio rather than its logarithm
as in conventional cosmological collider signals. The frequency of this oscillatory pattern is determined by the sound speed and the chemical potential, but not the mass of the exchanged vector field. Consequently, detecting a parity-violating trispectrum provides an interesting detection channel for extra massive and spinning particles, as well as for reconstructing parameters governing inflationary fluctuations.

\vskip 10pt
This paper is organized as follows. In Sec.~\ref{sec: Chemical Potential in the EFT of Inflation}, we start by introducing the chemical potential operator and the couplings of the massive spinning field to the Goldstone boson of broken time translations in the language of the EFT of inflationary fluctuations. After analyzing perturbativity bounds, we move on to the non-local single-field EFT description of the theory in Sec.~\ref{sec: Non-Local EFT}, and estimate the size of non-Gaussianities therein. We then compute the resulting parity-odd trispectrum signal and analyze its noticeable features in Sec.~\ref{sec: Results}. Finally, we conclude in Sec.~\ref{sec: Conclusion}.

\section{Chemical Potential in the EFT of Inflation}
\label{sec: Chemical Potential in the EFT of Inflation}

In this section, we construct the theory that couples the Goldstone boson of broken time translations $\pi$ to a massive vector field $\sigma_\mu$ with a chemical potential. We start by reviewing the EFT of inflationary fluctuations (see the original papers \cite{Cheung:2007st, Senatore:2010wk} or the review \cite{Piazza:2013coa} for more details). Then, we introduce the quadratic action for $\sigma_\mu$, highlighting important properties that follow from the chemical potential. Finally, we establish the leading interactions between $\pi$ and $\sigma_\mu$ leading to parity violation in cosmological correlators, and derive perturbativity bounds.

\subsection{Goldstone Boson Action}

During inflation, the vacuum expectation value of the background inflaton field breaks time-translation invariance. As such, inflation can be seen as a process of spontaneous symmetry breaking that gives rise to a Goldstone boson $\pi$ describing scalar fluctuations. In the unitary gauge, in which fluctuations 
of the clock field are absorbed in the metric, the most general Lagrangian for $\pi$ is constructed out of geometrical objects that are invariant under spatial diffeomorphisms $x^i \rightarrow x^i + \xi^i(\bm{x}, t)$, namely $g^{00}$, $\partial_\mu g^{00}$, the extrinsic curvature $K_{\mu\nu}$, $\nabla_\mu$, $t$, and the Riemann tensor $R_{\mu\nu\rho\sigma}$ \cite{Cheung:2007st}. Organized as a derivative expansion, the leading terms in the action can be written 
\begin{equation}
	\label{eq: unitary gauge action}
	\begin{aligned}
		S_\pi &= \int \d^4x \sqrt{-g} \left[\frac{1}{2}\Mpl^2 R + \Mpl^2 \dot{H} g^{00} \right.\\
		&\left.- \Mpl^2(3H^2 + \dot{H}) + \frac{M_2^4}{2} \left(\delta g^{00}\right)^2 + \cdots\right]\,,
	\end{aligned}
\end{equation}
where $\delta g^{00} = g^{00} + 1$, $M_2$ is a mass scale that we take to be constant, and ellipses denote higher-order terms. To ensure that the action starts quadratic in the fluctuations, the first two operators' coefficients were adjusted to eliminate tadpoles. The Goldstone boson $\pi$ associated with the spontaneous breaking of time-translation symmetry can be introduced after performing a spacetime-dependent time diffeomorphism $t\rightarrow t + \pi(\bm{x}, t)$, restoring full diffeomorphism invariance to all order of the theory. Typically, the resulting action is intricate as it mixes metric fluctuations with the Goldstone mode. However, for most applications of interest, metric fluctuations can be neglected by taking the so-called decoupling limit $\Mpl \rightarrow \infty$ and $\dot{H}\rightarrow 0$ while keeping the product $\Mpl^2\dot{H}$ fixed. In this regime, one can evaluate the Goldstone boson action in the unperturbed metric, and the transformation reduces to $\delta g^{00} \rightarrow -2\dot{\pi} - \dot{\pi}^2 + (\partial_i \pi)^2/a^2$. The action (\ref{eq: unitary gauge action}) becomes\footnote{We have only depicted operators that are fixed by the speed of sound to highlight the importance of symmetries, and therefore we have omitted additional self-interactions coming e.g.~from $(\delta g^{00})^3$.} 
\begin{equation}
	\label{eq: Goldstone boson action}
	\begin{aligned}
		S_\pi &= \int \d t\d^3x a^3 \left[ \frac{1}{2} \dot{\pi}_c^2 - \frac{c_s^2}{2} \frac{(\partial_i \pi_c)^2}{a^2} \right.\\
		&\left.+ \frac{c_s^{3/2}}{2f_\pi^2}(c_s^2-1)\dot{\pi}_c\frac{(\partial_i \pi_c)^2}{a^2} + \frac{c_s^3}{8f_\pi^4}(c_s^2 - 1) \frac{(\partial_i \pi_c)^4}{a^4} +\cdots \right]\,,
	\end{aligned}
\end{equation}
where 
$\pi_c= c_s^{-3/2} f_\pi^2\pi$ is the canonically normalized Goldstone boson, $f_\pi^4\equiv 2c_s M_{\text{pl}}^2 |\dot{H}|$ is the symmetry-breaking scale and $c_s^{-2} = 1 - 2M_2^4/\Mpl^2|\dot{H}|$ is the speed of sound of $\pi_c$. The scale $f_\pi$ is related to the measured amplitude of the dimensionless power spectrum $(2\pi)^2 \Delta_\zeta^2 = (H/f_\pi)^4$ and the comoving curvature perturbation $\zeta$ is linked to the Goldstone boson by $\zeta = -H c_s^{3/2}f_\pi^{-2}\pi_c + \mathcal{O}(\pi_c^2)$. Importantly, as a consequence of the Goldstone boson non-linearly realizing Lorentz symmetry, a reduced speed of sound $c_s\ll 1$ generates large parity-even non-Gaussianities $f_{\text{NL}} \sim 1/c_s^2$ and $\tau_{\text{NL}} \sim 1/c_s^4$. Therefore, looking at parity-violating contributions to non-Gaussianities is free from self-interaction signals. 

\subsection{Spin-1 Chemical Potential}

Let us now introduce a massive vector field $\sigma_\mu$ with a chemical potential that breaks parity, hence leaving distinctive signatures in the parity-odd trispectrum. 

\vskip 4pt
\textbf{Unitary gauge.} In the unitary gauge, we take the action for the spin-1 massive field to be of Proca type supplemented by a Chern-Simons term
\begin{equation}
	\label{eq: action for sigma}
	\begin{aligned}
		S_\sigma = \int \d^4x \sqrt{-g} \left[ -\frac{1}{4} F_{\mu\nu}^2 -  \frac{m^2}{2} \sigma_\mu^2 + \frac{\kappa t}{4} F_{\mu\nu}\tilde{F}^{\mu\nu} \right]\,,
	\end{aligned}
\end{equation}
where $\tilde{F}^{\mu\nu} = \mathcal{E}^{\mu\nu\alpha\beta}F_{\alpha\beta}$ with $\mathcal{E}^{\mu\nu\alpha\beta} = \frac{\epsilon^{\mu\nu\alpha\beta}}{\sqrt{-g}}$ being the Levi-Civita tensor density, and $\kappa$ is the chemical potential. It is clear that all the terms in the action are invariant under spatial diffeomorphisms. After reintroducing the Goldstone boson, that transforms as $\pi \rightarrow \pi - \xi^0(\bm{x}, t)$ under a time-diffeomorphism $t \rightarrow t + \xi^0(\bm{x}, t)$, the specific coupling of $\sigma_\mu$ to the time foliation of spacetime prevents the last term in (\ref{eq: action for sigma}) from being a total derivative. This term respects the shift symmetry for the Goldstone boson.

\vskip 4pt
From a model-building perspective, the chemical potential naturally arises from the dimension-5 coupling $\phi F \tilde{F}$ between
the inflaton field $\phi$ and the spin-1 field $\sigma_\mu$ \cite{Anber:2009ua,Barnaby:2010vf,Sorbo:2011rz,Barnaby:2011qe, Pajer:2013fsa}. This coupling respects the shift symmetry of the scalar field $\phi\rightarrow \phi + c$, since the constant $c$ contributes to a total derivative. With the inflaton acquiring its slowly rolling vacuum expectation value, i.e.~$\phi = \phi_0 + \dot{\phi}_0t + \cdots$, the chemical potential term is identified as $\kappa = \dot{\phi}_0/\Lambda_\kappa$ where $\Lambda_\kappa$ is some higher-energy cutoff. Higher-order terms are slow-roll suppressed. From an EFT point of view, the chemical potential is a direct consequence of broken time diffeomorphism invariance by the background inflaton field. 

\vskip 4pt
\textbf{Equation of motion.} Varying the action (\ref{eq: action for sigma}) yields the equation of motion
\begin{equation}
	\label{eq: EOM sigma}
	\Box \sigma^\nu - \nabla_\mu \nabla^\nu \sigma^\mu - m^2 \sigma^\nu - \kappa \mathcal{E}^{0\nu\alpha\beta} F_{\alpha\beta} = 0\,,
\end{equation} 
where $\Box \equiv g^{\mu\nu}\nabla_\mu \nabla_\nu = -\partial_t^2 - 3H\partial_t + \partial_i^2/a^2$ denotes the d'Alembert operator on four-dimensional de Sitter spacetime. The time-like component of the field $\sigma_\mu$ is solved in terms of the spatial components using the constraint equation $\nabla^\mu \sigma_\mu = 0$, which is found after taking the divergence of the equation of motion. Substituting the constraint back into (\ref{eq: EOM sigma}) gives the following on-shell equation of motion
\begin{equation}
	\left[\Box - (m^2 + 3H^2)\right]\sigma^\nu = 2\kappa \mathcal{E}^{0\nu\alpha\beta}\nabla_\alpha \sigma_\beta\,.
\end{equation}
Being massive, the field $\sigma_\mu$ has three degrees of freedom, a longitudinal mode and two transverse modes. Decomposing the spatial components in 3-dimensional Fourier space $\sigma_i(\bm{x}, \tau) = \int \frac{\d^3k}{(2\pi)^3}\, e^{i \bm{k}\cdot \bm{x}}\, \sigma_{i, \bm{k}}(\tau)$ into helicity eigenstates, one can write
\begin{equation}
	\sigma_{i, \bm{k}}(\tau) = \sum_{\lambda=0, \pm} \varepsilon_i^{\lambda} \left[\sigma_{\lambda, k}(\tau) \, a_{\bm{k}}^\lambda + \sigma_{\lambda, k}^*(\tau) \, a_{-\bm{k}}^{\lambda \dagger}\right]\,,
\end{equation}
where we have used conformal time $\tau$ defined as $\mathrm{d}\tau \equiv \mathrm{d}t/a$, and where $\sigma_{\lambda, k}(\tau)$ are the helicity eigenstates, with $\lambda = 0$ being the longitudinal mode and $\lambda = \pm$ denoting the transverse modes. In terms of these mode functions, the equations of motion for different helicities decouple from each other. As we are interested in parity violation, and the chemical potential only affects the transverse modes, we only consider the transverse modes equation of motion that reads
\begin{equation}
	\sigma_{\pm, k}'' + \left[k^2 \pm 2ak\kappa  + a^2m^2\right]\sigma_{\pm, k} = 0\,,
 \label{eq:eom}
\end{equation} 
where a prime denotes a derivative with respect to conformal time $\tau$. 

\vskip 4pt
\textbf{Dynamics of transverse modes.} The mode functions of each helicity eigenstate $\sigma_{\lambda, k}(\tau)$ can be solved analytically in terms of Whittaker $W$-functions, see App.~\ref{sec: ExactUVApp}. However, concentrating on transverse modes, it is more instructive to directly read the physics from the equation of motion \eqref{eq:eom}. In particular, in the flat-space limit $H \rightarrow 0$, neglecting Hubble friction for simplicity, the equation of motion can be solved with plane waves, leading to the following dispersion relation
\begin{equation}
    \label{eq: spin-1 dispersion relation}
    \omega_\pm^2(\kp) = (\kp\pm \kappa)^2 + m^2 - \kappa^2\,,
\end{equation}
where $\kp = k/a$ is the physical momentum. The effect of the chemical potential appears as a constant shift of $\kp$, or equivalently as a time-dependent mass term owing to the redshift of the physical momentum. This phenomenon assists the spontaneous particle production of $\sigma_\pm$ by the exponentially expanding spacetime, enhancing the cosmological collider signal \cite{Wang:2019gbi, Sou:2021juh}. However, the latter effect does not manifest itself in the single-field EFT treatment of interest in this work. Instead, what is of direct relevance for our purpose is simply the \textit{asymmetry} between the different helicities, as it is this parity violation that gets imprinted in the four-point correlator of the Goldstone boson.
Considering $\kappa >0$ without loss of generality, the dispersion relation \eqref{eq: spin-1 dispersion relation} shows that for $\kappa>m$, each $\sigma_-$ mode experiences a transient tachyonic instability in the time interval with $ \omega_-^2(\kp)<0$. The energy carried by these modes can lead to a strong backreaction on the inflationary background and, in what follows, we restrict our analysis to the situation with $\kappa \leq m$. In this regime, the behavior of each helicity mode is qualitatively similar, with modes beginning to decay around the mass-shell horizon $\kp/H=|k\tau|\sim m/H$. Nonetheless, the chemical potential does affect differently the two helicities, with important consequences as we will see.

\subsection{Mixing Interactions}

Let us now construct the leading couplings between the spin-1 field $\sigma_\mu$ and the Goldstone boson $\pi$. 

\vskip 4pt
In the unitary gauge, Lorentz invariant operators constructed out of the spin-1 field, e.g.~$\sigma_\mu \sigma^\mu$, are of course invariant under all diffeomorphisms and therefore are invariant under spatial ones. However, they do not lead to mixing interactions with the Goldstone boson. The time-like component $\sigma^0$ is a scalar under spatial diffeomorphisms, hence it constitutes a building block to construct mixing interactions. Performing a spacetime-dependent time diffeomorphism, reintroducing the Goldstone boson, leads to 
\begin{equation}
	\begin{aligned}
		\sigma^0(\bm{x}, t) \rightarrow \tilde{\sigma}^0(\tilde{\bm{x}}, \tilde{t}) &= \partial_\mu (t + \pi) \,\sigma^\mu(\bm{x}, t) \\
		&= \left[\delta_\mu^0 + \partial_\mu \pi\right]\sigma^\mu(\bm{x}, t)\,.
	\end{aligned}
\end{equation}
In order to generate a parity-odd trispectrum via the (single) exchange of $\sigma_\mu$, one needs to determine mixing interactions of the schematic form $\sim \pi^2\sigma$, compatible with the symmetry and with an odd number of spatial derivatives. There are two ways to generate such interactions: (i) couple the time-like component $\sigma^0$ to index-free building blocks invariant under spatial diffeomorphisms, and (ii) couple $\sigma_\mu$ with building blocks carrying a free (spacetime) index. Up to dimension seven, the corresponding lowest-dimension cubic interactions are therefore given by the following operators
\begin{align}
	\nonumber S_{\pi\sigma} = \int \d^4x\sqrt{-g} &\Big[\omega_1^3 \delta g^{00} \sigma^0 + \omega_2 \nabla_\mu \delta g^{00} \delta K_\nu^\mu \sigma^\nu \\
	&+ \omega_3 n^\mu\nabla_\mu\delta g^{00}\nabla_\nu \delta g^{00}\sigma^\nu\Big]\,,
\end{align}
where $\omega_{1, 2, 3}$ are constant mass-dimension couplings, and $\delta K_\mu^\nu = K_{\mu\nu} - a^2 H h_{\mu\nu}$ with $h_{\mu\nu}$ the induced spatial metric. Reintroducing the Goldstone boson leads to 
\begin{equation}
	\label{eq: mixing action with pi-sigma}
	\begin{aligned}
		S_{\pi\sigma} = \int &\d t\d^3x a^3 \Bigg[\rho a^{-2} \partial_i \pi_c\sigma_i - \frac{a^{-2}}{\tilde{\Lambda}}\dot{\pi}_c \partial_i \pi_c \sigma_i \\
		&+ \frac{a^{-4}}{\Lambda^3} \partial_j \dot{\pi}_c \partial_i \partial_j \pi_c \sigma_i+\frac{a^{-2}}{\bar{\Lambda}^3}\ddot{\pi}_c\partial_i\dot{\pi}_c\sigma_i\Bigg]\,,
	\end{aligned}
\end{equation}
where the couplings are defined by $\rho \equiv 2c_s^{3/2} \omega_1^3/f_\pi^2$, $\tilde{\Lambda}^{-1} \equiv c_s^{3/2} \rho/f_\pi^2$, $\Lambda^{-3} \equiv 2 c_s^3 \omega_2/f_\pi^4$, and $\bar{\Lambda}^{-3}\equiv 4 c_s^3\omega_3/f_\pi^4$. Here we have kept cubic terms in the fields relevant for the parity-odd trispectrum. Note that the size of the cubic interaction
$\dot{\pi}_c \partial_i \pi_c \sigma_i$ is completely fixed by the size of the quadratic mixing $\partial_i \pi_c \sigma_i$. This is a consequence of the Goldstone boson non-linearly realizing time diffeomorphism, as both interactions are governed by the same coupling $\omega_1$.\footnote{Notice that we could have considered manifestly parity-violating interactions by contracting spatial derivatives with $\epsilon_{ijk}$. The leading operator in this case would be a dimension-8 operator and the corresponding exchange diagram would typically give a smaller signal compared to the one we compute using parity-even vertices only.} We will see later that the two operators with coupling proportional to $\omega_1$ are tightly constrained by mixing perturbativity, leading to undetectably small parity-violating non-Gaussianities. However, the two dimension-7 operators originating from $\omega_2$ and $\omega_3$ can lead to sizable signals. Note that by putting the Goldstone boson on shell, namely using its equation of motion, and observing that $\partial_i (\dot{\pi}^2_c)$ does not couple to the transverse mode of the vector field, the cubic interaction governed by the coupling $\omega_3$ is of the same form as that fixed by $\omega_2$, albeit with a different spatial gradient structure.

\subsection{Perturbativity Bounds}
\label{sec: Perturbativity Bounds}

The introduced couplings in the mixing action (\ref{eq: mixing action with pi-sigma}) are free parameters of the theory. 
We will derive bounds on them by requiring that the dimension-3 quadratic mixing term can be treated perturbatively, and that the dimension-5 and -7 interactions are suppressed by scales larger than $H$, ensuring weak coupling of fluctuations.
The corresponding bounds on $\rho$, $\Lambda$ and $\bar{\Lambda}$ will eventually determine how large the corresponding non-Gaussianities can be.

\vskip 4pt
\textbf{Quadratic theory.} By dimensional analysis, we estimate the size of the various terms in the total action and require that the mixing interactions be smaller than the smallest kinetic term. As the terms are time dependent (momenta redshift due to the expanding spacetime), we will eventually evaluate all quantities at the energy set by the Hubble scale $H$. Assuming weak quadratic mixing, the dispersion relations for the Goldstone boson and the transverse modes of the massive spin-1 field are $\omega_\pi = c_s \kp$ and $\omega_\sigma^2 = \kp^2 \pm 2\kappa \kp + m^2$, respectively. A consequence of the Goldstone boson having a reduced speed of sound, hence breaking de Sitter boosts, is that time and space cannot be treated on the same footing. One needs to keep track of how terms in the action scale with energy \textit{and} (physical) momentum. Taking the physical momentum $\kp$ as the reference, both dispersion relations imply $c_s^2 \omega_\sigma^2 = \omega^2 \pm 2c_s \kappa \omega + c_s^2 m^2$, where we have denoted $\omega_\pi \equiv \omega$, the energy scale probed during inflation. The Goldstone boson kinetic term then scales as $\sim \omega^2 \pi_c$, and that of the spin-1 field reads $\sim (\omega^2/c_s^2)\, \sigma^2\g(\omega)$, where we have defined $\g(\omega) = 1 \pm 2\alpha_\kappa(\omega) + \alpha_m^2(\omega)$ in terms of the dimensionless energy-dependent quantities $\alpha_\kappa(\omega) \equiv c_s \kappa/\omega$ and $\alpha_m(\omega) \equiv c_s m/\omega$. As we will see in the following, the regime of interest, where physical effects of a reduced speed of sound become important, is characterized by $\alpha_\kappa\ll 1$ and $\alpha_m\ll 1$, hence $\g\sim 1$, where the parameters are evaluated at the Hubble scale $\omega=H$. We nevertheless keep these terms for consistency. Note that the mass term of the additional field and the chemical potential term are already taken into account in the dispersion relation. For a reduced sound speed, they do not significantly contribute to the quadratic action for both fields and can be consistently neglected. We now need to determine the typical size of the fluctuations $\pi_c$ and $\sigma$. By dimensional analysis, requiring $\int \d t\d^3x \dot{\pi}_c^2\sim 1$, we obtain $\pi_c\sim c_s^{-3/2}\omega$. Similarly, we obtain $\sigma \sim \omega/c_s \,\g^{-1/4}(\omega)$. Combining these scalings with those of the free quadratic terms in the action, the Goldstone boson kinetic term scales as $\sim c_s^{-3} \omega^4$ and the kinetic term of the spin-1 field scales as $\sim c_s^{-4}\omega^4\g^{1/2}(\omega)$. It becomes clear that the Goldstone boson kinetic term is smaller than that of the massive field for $c_s \ll 1$. 

\vskip 4pt
\textbf{Mixing interactions.} Using the same dimensional analysis arguments, the mixing interactions scale as 
\begin{equation}
	\begin{aligned}
		\rho \,\partial_i \pi_c \sigma_i &\sim c_s^{-1/2} \rho \,\frac{\omega^3}{c_s^3} \,\g^{-1/4}(\omega)\,, \\
		\frac{1}{\tilde{\Lambda}} \dot{\pi}_c \partial_i \pi_c \sigma_i &\sim c_s^{-7} \frac{\omega^5}{\tilde{\Lambda}} \,\g^{-1/4}(\omega)\,, \\
		\frac{1}{\Lambda^3} \partial_j \dot{\pi}_c \partial_i \partial_j \pi_c \sigma_i &\sim c_s^{-7} \frac{\omega^7}{\Lambda^3}\,\g^{-1/4}(\omega)\,,\\
		\frac{1}{\bar{\Lambda}^3} \ddot{\pi}_c \partial_i \dot{\pi}_c \sigma_i &\sim c_s^{-5} \frac{\omega^7}{\bar\Lambda^3}\,\g^{-1/4}(\omega)\,.
	\end{aligned}
\end{equation}
Let us now require that these scalings should be smaller than the Goldstone kinetic term which is of order $c_s^{-3}\omega^4$, and evaluate the above expressions at $\omega=H$. For the quadratic coupling $\rho$, which also fixes the size of the cubic interactions $\dot{\pi}_c \partial_i \pi_c \sigma_i$, one obtains 
\begin{equation}
	\label{eq: perturbativity bound rho}
	\frac{\rho}{H} \lesssim c_s^{1/2} \g^{1/4}\,, \hspace*{0.2cm} \text{and}\hspace*{0.2cm} \frac{\rho}{H} \lesssim c_s^{5/2} \frac{\g^{1/4}}{2\pi \Delta_\zeta}\,.
\end{equation}
Within our regime of interest, $c_s\ll 1$ and $\g\sim 1$, the first inequality gives the most stringent bound and is taken to be the perturbative criterion.\footnote{In the case of a vanishing chemical potential, a more stringent bound on the quadratic mixing strength $\rho$ can be found by requiring that the spin-1 field propagates subluminally \cite{Lee:2016vti}. This bound enforces the coupling $\rho$ to be almost vanishing if the longitudinal mode of the massive vector field propagates almost relativistically, and therefore puts severe theoretical constraints on the size of non-Gaussianities. In practice, this leads to undetectable parity-even and parity-odd non-Gaussian signals generated by mixings fixed by the coupling $\rho$.} Performing the same procedure for the remaining interactions gives
\begin{equation}
	\label{eq: perturbativity bound Lambda}
	\frac{H}{\Lambda} \lesssim c_s^{4/3} \g^{1/12}\,, \hspace*{0.2cm} \text{and}\hspace*{0.2cm} \frac{H}{\bar\Lambda} \lesssim c_s^{2/3} \g^{1/12}~.
\end{equation}
We will see in the following that both $\partial_j \dot{\pi}_c \partial_i \partial_j \pi_c \sigma_i$ and $\ddot{\pi}_c\partial_i\dot{\pi}_c\sigma_i$ can lead to a large parity-odd trispectra. In fact, the generated trispectrum shapes are similar, and we will mainly focus on the former one without loss of generality. Note that the way the spin-1 field couples to the time foliation through the chemical potential term $\kappa t F_{\mu\nu} \tilde{F}^{\mu\nu}$ introduces an additional mixing of the form $\sim \pi \sigma^2$. The same scaling analysis as previously gives the bound $\kappa/H \lesssim  c_s\g^{-1/2}/(2\pi  \Delta_\zeta)$ 
on the chemical potential. However, this mixing contributes to the (double-exchange) bispectrum and the (triple-exchange) trispectrum, and typically leads to no observable non-Gaussianity. We therefore discard this mixing interaction from our analysis.

\vskip 4pt
\textbf{Perturbative unitarity.} The UV cutoff of the Goldstone boson sector critically approaches the Hubble scale when the speed of sound is reduced. Indeed, one can show that for energies above \cite{Cheung:2007st, Baumann:2011su, Baumann:2014cja}
\begin{equation}
	\Lambda_\star \approx 100 c_s H\,, 
\end{equation}
the perturbative description breaks down as self-interactions become comparable to the kinetic term of the Goldstone boson.\footnote{Note that the previously derived bounds in Eqs.~(\ref{eq: perturbativity bound rho})-(\ref{eq: perturbativity bound Lambda}) of course also define strong coupling scales associated with mixing interactions. The strong coupling scale of the full theory, including all sectors, should be taken to be the most stringent one.} In flat-space particle physics, the massive field $\sigma_\mu$ cannot be heavier than this UV cutoff for theoretical consistency. Indeed, its subsequent decay into Goldstone bosons would imply, by energy conservation, that the low-energy degree of freedom carries energy above $\Lambda_\star$, leading to the breakdown of unitarity. The equivalent process in cosmology is the on-shell production of heavy fields which leads to the standard cosmological collider signature. At the level of the two-field description of fluctuations, the massive field therefore cannot be too heavy, i.e.~$m\lesssim \Lambda_\star$. However, in what follows, we will precisely take into account the effect of the field $\sigma_\mu$ by integrating it out from the spectrum, leading to a single-field low-energy description. Of course, this description misses the non-perturbative particle production, but accurately captures all other effects emerging from a reduced sound speed and the presence of a chemical potential. 

\section{Non-Local EFT}
\label{sec: Non-Local EFT}

Having introduced the considered theory, let us now delve into the analysis of the underlying physics. We will work in the so-called low-speed collider regime, where $ m\lesssim H/c_s$, and qualitatively new phenomena arise, see \cite{Jazayeri:2022kjy,Jazayeri:2023xcj} for details. 
In brief, by momentum conservation, the massive field typically has a comparable wavelength as the Goldstone boson. However, the Goldstone mode has a much smaller sound horizon, which means that the massive mode 
can be considered as effectively massless when the Goldstone freezes. The interaction between Goldstone modes outside the sound horizon and massive modes in the quantum vacuum leads to striking non-Gaussian signatures, generically referred to as low-speed collider resonances.

\vskip 4pt
Stemming from a reduced Goldstone sound speed, a notable characteristic of the low-speed collider is that the vector field response in the bulk of the inflationary spacetime can be treated as instantaneous. This results in a non-local single-field EFT description that 
captures well the underlying physics. This non-locality, although mild as we will see, turns out to be critical for bringing out the parity violation to the final observables.

\subsection{Emergent Non-Locality}

We start by recalling the UV theory for the Goldstone boson and the transverse component of the vector field. Explicitly writing the mixing of the vector field with the Goldstone boson through its coupling to a current $J_i(\pi_c)$, the action reads
\begin{align}
	\label{eq: UVLag}
	\nonumber &S_{\text{UV}}=\int \d t\d^3 x a^3\Bigg[\frac{1}{2}\dot{\pi}_c^2-\frac{c_s^2}{2}\frac{(\partial_i\pi_c)^2}{a^2}\\
	&-\frac{1}{2a^2}\sigma_i\left[\delta_{ij}(\partial_t^2 + H \partial_t) + \mathcal{D}_{ij}\right]\sigma_j+ \frac{\sigma_i}{a^2} J_i(\pi_c)\Bigg]\,,
\end{align}
where we have introduced the spatial component of the quadratic kinetic operator of the massive spinning field
\begin{align}
	\D_{ij}\equiv \left(-\frac{\partial_i^2}{a^2} + m^2\right)\delta_{ij} - 2\kappa \epsilon_{ijl}\frac{\partial_l}{a}\,,
\end{align}
that can be identified as the effective frequency of the vector field in Fourier space, and the Goldstone currents (up to a total derivative) from the operators in (\ref{eq: mixing action with pi-sigma})
\begin{align}
	J_i(\pi_c)\equiv -\frac{1}{\tilde\Lambda}\dot{\pi}_c\partial_i\pi_c+ \frac{a^{-2}}{\Lambda^3}\partial_j\dot{\pi}_c\partial_i\partial_j\pi_c+\frac{1}{\bar{\Lambda}^3}\ddot{\pi}_c\partial_i\dot{\pi}_c\,.
\end{align}
The Gaussian form of the vector sector in the UV Lagrangian (\ref{eq: UVLag}) allows one to directly integrate out $\sigma_i$ and obtain an effective theory, if the massive vector field is not excited on-shell. The effective theory is necessarily non-local in time and space, and can be formally written as
\begin{align}
	\nonumber S_{\text{EFT}}&=\int \d t\d^3 x a^3 \Bigg[\frac{1}{2}\dot{\pi}_c^2-\frac{c_s^2}{2}\frac{(\partial_i\pi_c)^2}{a^2}\\  
	&+\frac{1}{2a^2}J_i(\pi_c)\left[(\partial_t^2 + H \partial_t) + \mathcal{D}\right]^{-1}_{ij}J_j(\pi_c)\Bigg]\,.
\end{align}
Since the vector response is much faster than the relaxation of the Goldstone, 
one can expand the second line in powers of $-(\partial_t^2 + H \partial_t)/\D\sim \omega_\pi^2/\omega_\sigma^2\sim c_s^2\ll 1$. The fully non-local differential operator becomes effectively non-local in space only, and reads
\begin{align}
	\label{eq: timeDerivativeExpansion}
	\left[(\partial_t^2 + H \partial_t)+\D\right]^{-1}_{ij}=\D^{-1}_{il}\sum_{n=0}^{\infty}\left[-(\partial_t^2 + H \partial_t)\,\D^{-1}\right]^n_{lj}\,.
\end{align}
This allows one to construct a non-local single-field EFT organized as a \textit{time-derivative expansion}, whose first leading terms are
\begin{align}
	\label{eq: IRLag}
	\nonumber S_{\text{EFT}}&=\int \d t\d^3 x a^3\Bigg[\frac{1}{2}\dot{\pi}_c^2-\frac{c_s^2}{2}\frac{(\partial_i\pi_c)^2}{a^2}
	\\
	\nonumber&+\frac{1}{2a^2}J_i(\pi_c)\D^{-1}_{ij}J_j(\pi_c)\\
	&-\frac{1}{2a^2}J_i(\pi_c)\D^{-1}_{il}(\partial_t^2 + H \partial_t) \,\D^{-1}_{lj} J_j(\pi_c)
	\\
	\nonumber&+\mathcal{O}(\partial_t^4)\Bigg]\,.
\end{align}
Here, only the Leading-Order (LO) and the Next-to-Leading-Order (NLO) non-local EFT operator are spelled out explicitly, corresponding to the second and third line, respectively. The non-local differential operator $\D^{-1}$ can be projected onto parity-even and parity-odd parts
\begin{align}
	\D^{-1}_{ij}=[\D^{-1}_{\mathrm{PE}}]_{ij}+[\D^{-1}_{\mathrm{PO}}]_{ij}\,,
\end{align}
with
\begin{align}
	\label{eq: non-local PE/PO operators}
	[\D^{-1}_{\mathrm{PE}}]_{ij}&\equiv\frac{(-\partial_i^2/a^2+m^2)\left[\delta_{ij}+\frac{4\kappa^2\partial_i\partial_j/a^2}{(-\partial_i^2/a^2+m^2)^2}\right]}{(-\partial_i^2/a^2+m^2)^2+4\kappa^2\partial_i^2/a^2}\,,\\
	[\D^{-1}_{\mathrm{PO}}]_{ij}&\equiv\frac{2\kappa \epsilon_{ijl}\partial_l/a}{(-\partial_i^2/a^2+m^2)^2+4\kappa^2\partial_i^2/a^2}\,.
\end{align}
Before explicitly computing the trispectrum, we make a few comments on the non-local single-field theory (\ref{eq: IRLag}).

\begin{enumerate}
	\item[$\bullet$] The appearance of inverse spatial derivatives in $\D^{-1}_{ij}$ implies that the action actually contains terms evaluated at distinct spatial positions, i.e.~that the action is spatially non-local. This does not signal any pathological behavior, as such form of non-locality is mild: the effect of the action at distance that this describes is confined within the Compton wavelength of the massive spin-1 field, analogous to the non-local Yukawa force after integrating out the pions in nuclear theory \cite{Jazayeri:2023xcj}.
  For the same reason, correlators generated by (contact) interactions in the theory (\ref{eq: IRLag}) satisfy the Manifest Locality Test (MLT) \cite{Jazayeri:2021fvk}. Indeed, $\D^{-1}_{ij}$, once expressed in Fourier space, has no singularity in the long wavelength limit $k  \to 0$, due to the displacement of its pole away from $\partial_i^2=0$ by the presence of the finite mass.

	\item[$\bullet$] The formal series in Eq.~(\ref{eq: timeDerivativeExpansion}) is actually divergent. For $\D$ being a rational function of the scale factor, the time derivatives bring a factorial contribution $\partial_t^{2n}\sim (2n)!$, which always dominates over the EFT expansion $\D^{-2n}$ in the large-order limit. Such a divergence is not a surprise, since we expect non-perturbative effects that are not captured by the $\D^{-1}$ expansion to kick in at large orders, and indeed cosmological collider signals coming from on-shell particle propagation are \textit{not} captured by the non-local EFT.\footnote{Another perspective on this can be gained by examining the time-dependence of the vector field equation of motion. Namely, the time dependence in $\D$---hidden in the physical momenta $\kp=k/a(t)$---leads to $[\partial_t^2 + H\partial_t,\D^{-1}]\neq 0$, which is the cause of the $(2n)!$ behavior at large orders. Physically this corresponds to the Stokes phenomenon for time-dependent equation of motions with particle production, see e.g.~\cite{Sou:2021juh}.} For our purpose, apart from internal ultra-soft limits of correlators $k_\textrm{L}/k_\textrm{S} \lesssim c_s$ where this effective description breaks down for $m \sim H$ \cite{Jazayeri:2022kjy,Jazayeri:2023xcj}, it suffices to truncate the time-derivative expansion (\ref{eq: timeDerivativeExpansion}) at LO or NLO. As we will see in what follows, the noteworthy aspect of the non-local EFT is its remarkable capability to provide a highly accurate trispectrum from regular to mildly-soft kinematic configurations.
	
	\item[$\bullet$] Note that the non-local EFT contains a single massless degree of freedom with a Bunch-Davies vacuum. It is also invariant under a global dilation transformation $(t,\bm{x})\to(t +\Delta t,\bm{x}e^{-H\Delta t})$, which is equivalent to scale invariance. Additionally, the whole tower of interactions in Eq.~(\ref{eq: IRLag}) generates IR-finite correlators. Thus the non-local EFT satisfies all the assumptions of the no-go theorem on parity violation except the implicit locality requirement \cite{Liu:2019fag, Cabass:2022rhr}. This \textit{emergent} non-locality is derived from the fact that the UV theory contains an extra massive spinning field. 
\end{enumerate}

From the UV picture, it is clear that the origin of parity-violating observables stems from the intrinsic chemical-potential induced parity-odd dynamics of the vector field. This parity-odd dynamics subsequently propagates to the visible sector through parity-even mixing interactions. From the non-local single-field EFT, after inspecting the structure of the non-local parity-odd operator $\mathcal{D}^{-1}_{\text{PO}}$ in Eq.~(\ref{eq: non-local PE/PO operators}), it becomes manifest that the theory leads to a non-vanishing parity-violating trispectrum even without any explicit reference to the UV massive field.

\subsection{Size of Parity-Odd Signals}

The (non-local) single-field theory (\ref{eq: IRLag}) provides a handy way for estimating the size of the parity-odd trispectrum, denoted $\tau_{\text{NL, PO}}$, generated by the exchange of a massive spinning field $\sigma_i$. The theory contains only one dynamical degree of freedom, the Goldstone boson, and a simple estimate of the parity-odd trispectrum is then given by $\tau_{\text{NL, PO}} \sim \frac{\mathcal{L}_{4, \text{PO}}}{\mathcal{L}_2} \Delta_\zeta^{-2}$, where all dimensionful quantities in the Lagrangian, which crucially depend on time, are evaluated at sound horizon crossing $c_s \kp \sim H$, or equivalently at the Hubble scale $\omega \sim H$.\footnote{This estimate does not take into account the interactions outside the sound horizon, characteristic of the low-speed collider physics, that give rise to additional dependencies on $\alpha_{m,\kappa}$, see Sec.~\ref{sec: Results}. However, it is sufficient for our purpose here to estimate the size of the signals.} The quadratic Lagrangian $\mathcal{L}_2$ is taken to be the Goldstone boson kinetic term that scales as $\sim c_s^{-3} \omega^4$, as seen in Sec.~\ref{sec: Perturbativity Bounds}. Evaluating the parity-odd non-local differential operator $\mathcal{D}^{-1}_{\text{PO}}$ in Eq.~(\ref{eq: non-local PE/PO operators}) at the relevant time scale leads to 
\begin{equation}
	\mathcal{D}^{-1}_{\text{PO}} \sim \left(\frac{c_s}{H}\right)^2 \, \frac{\alpha_\kappa}{(1 + \alpha_m^2)^2 + 4\alpha_\kappa^2}\,,
\end{equation}
where $\alpha_m = c_s m/H$ and $\alpha_\kappa = c_s \kappa/H$ are small in the regime of interest $c_s\ll 1$, also corresponding to the domain of validity of the non-local EFT. In the following, for simplicity, we will estimate the size of the parity-odd exchange trispectra involving the same two vertices, although the procedure analogously extends to all contributions. 

\vskip 4pt
\textbf{Dimension-5 operator.} After integrating out the massive spinning field in a non-local manner, the dimension-5 interaction in the Lagrangian reads $\mathcal{L}_4/a^3 = -\frac{a^{-2}}{\tilde{\Lambda}^2} \dot{\pi}_c \partial_i \pi_c [\mathcal{D}^{-1}_{\text{PO}}]_{ij} \dot{\pi}_c \partial_j \pi_c$ and the size of the parity-odd trispectrum is given by 
\begin{equation}
	\tau_{\text{NL, PO}}^{\dot{\pi}_c \partial_i \pi_c \sigma_i} \sim \left(\frac{\rho}{H}\right)^2\, \frac{\alpha_\kappa}{(1 + \alpha_m^2)^2 + 4\alpha_\kappa^2}\,,
\end{equation}
where we recall that the energy-scale $\tilde{\Lambda}$ is fixed by the quadratic mixing strength $\rho$ due to the non-linearly realized symmetry. Using the perturbativity bound (\ref{eq: perturbativity bound rho}), and in the regime of interest $\alpha_m, \alpha_\kappa \ll 1$ and $\gamma \sim 1$, we find $\tau_{\text{NL, PO}}^{\dot{\pi}_c \partial_i \pi_c \sigma_i} \lesssim c_s \alpha_\kappa$, i.e.~negligible parity-odd non-Gaussianities. Henceforth, we will discard the contribution from the dimension-5 operator in the following calculations.

\vskip 4pt
\textbf{Dimension-7 operators.} Remarkably, the considered higher-order operators can generate large parity-odd non-Gaussianities. Indeed, integrating out the field $\sigma_i$ for the interactions fixed by the scale $\Lambda$ leads to $\mathcal{L}_4/a^3 = \frac{a^{-6}}{\Lambda^6} \partial_l \dot{\pi}_c \partial_i \partial_l \pi_c [\mathcal{D}^{-1}_{\text{PO}}]_{ij}\partial_k \dot{\pi}_c \partial_j \partial_k \pi_c$. The size of the resulting parity-odd trispectrum reads
\begin{equation}
	\label{eq: tauNL dim 7}
	\tau_{\text{NL, PO}}^{\partial_j \dot{\pi}_c \partial_i \partial_j \pi_c \sigma_i} \sim c_s^{-7} \left(\frac{H}{\Lambda}\right)^6 \, \frac{\alpha_\kappa \, \Delta_\zeta^{-2}}{(1 + \alpha_m^2)^2 + 4\alpha_\kappa^2}\,.
\end{equation}
We will see in Sec.~\ref{sec: Results} that the actual size of the trispectrum presents an additional $c_s^4$ suppression. This results from a careful analysis of the dimensionless integral entering in the size of the trispectrum, and can be traced back to the residue at the pole of the non-local differential operator $\mathcal{D}^{-1}_{\text{PO}}$. Accordingly, using the perturbativity bound (\ref{eq: perturbativity bound Lambda}) gives $\tau_{\text{NL, PO}}^{\partial_j \dot{\pi}_c \partial_i \partial_j \pi_c \sigma_i} \lesssim c_s^5 \alpha_\kappa \Delta_\zeta^{-2} = \mathcal{O}(500)$ for a representative value $c_s = 0.1$. For comparison, the parity-even trispectrum has been constrained to be $\tau_{\text{NL, PE}} \lesssim 2800$ and $g_{\text{NL, PE}}\lesssim 10^4$~\cite{Planck:2013wtn, Planck:2019kim}. Let us stress that the estimated parity-odd signal is free from large contaminations coming from self-interactions of the Goldstone boson. Although it might seem counter-intuitive, as the signal is generated by higher-derivative interactions, the reason for the apparent sound-speed enhancement in Eq.~(\ref{eq: tauNL dim 7}) can be traced back to the large number of spatial derivatives $\partial_i/a \sim \kp \sim H/c_s$, which are boosted in the case of a reduced sound speed $c_s \ll 1$. However, we stress that this enhancement is dummy when applying perturbativity bounds on the couplings. In fact, these mixings can generate large parity-odd signals due to the $\Delta_\zeta^{-2}$ enhancement as these interactions are not tied to the strength of the quadratic mixing. Following the same procedure for the interaction fixed by the energy scale $\bar{\Lambda}$ leads to the same bound $\tau_{\text{NL, PO}}^{\ddot{\pi}_c \partial_i \dot{\pi}_c \sigma_i} \lesssim c_s^5 \alpha_\kappa \Delta_\zeta^{-2}$, which also gives rise to a potentially large amount of parity-odd non-Gaussianity. In what follows, we will focus on the interaction $\partial_j \dot{\pi}_c \partial_i \partial_j \pi_c \sigma_i$ fixed by $\Lambda$ as both dimension-7 operators lead to qualitatively similar shapes with comparable sizes.

\section{Parity-Violating Trispectrum}
\label{sec: Results}

In this section, we show explicitly how the effect of the chemical potential in our non-local EFT enables one to bypass the single-field no-go theorem for parity violation, stressing the singularity structure in the complex plane of the operator ${\cal D}^{-1}$, which results in simple analytical expressions for the parity-odd trispectrum. Then, we uncover interesting features of the kinematic dependence of this correlator, including a low-speed collider resonance and a new class of oscillations, drastically distinct from the conventional cosmological collider signal.

\subsection{Parity Violation as a Residue}

\begin{figure}[htbp]
	\centering
	\includegraphics[width=0.27\textwidth]{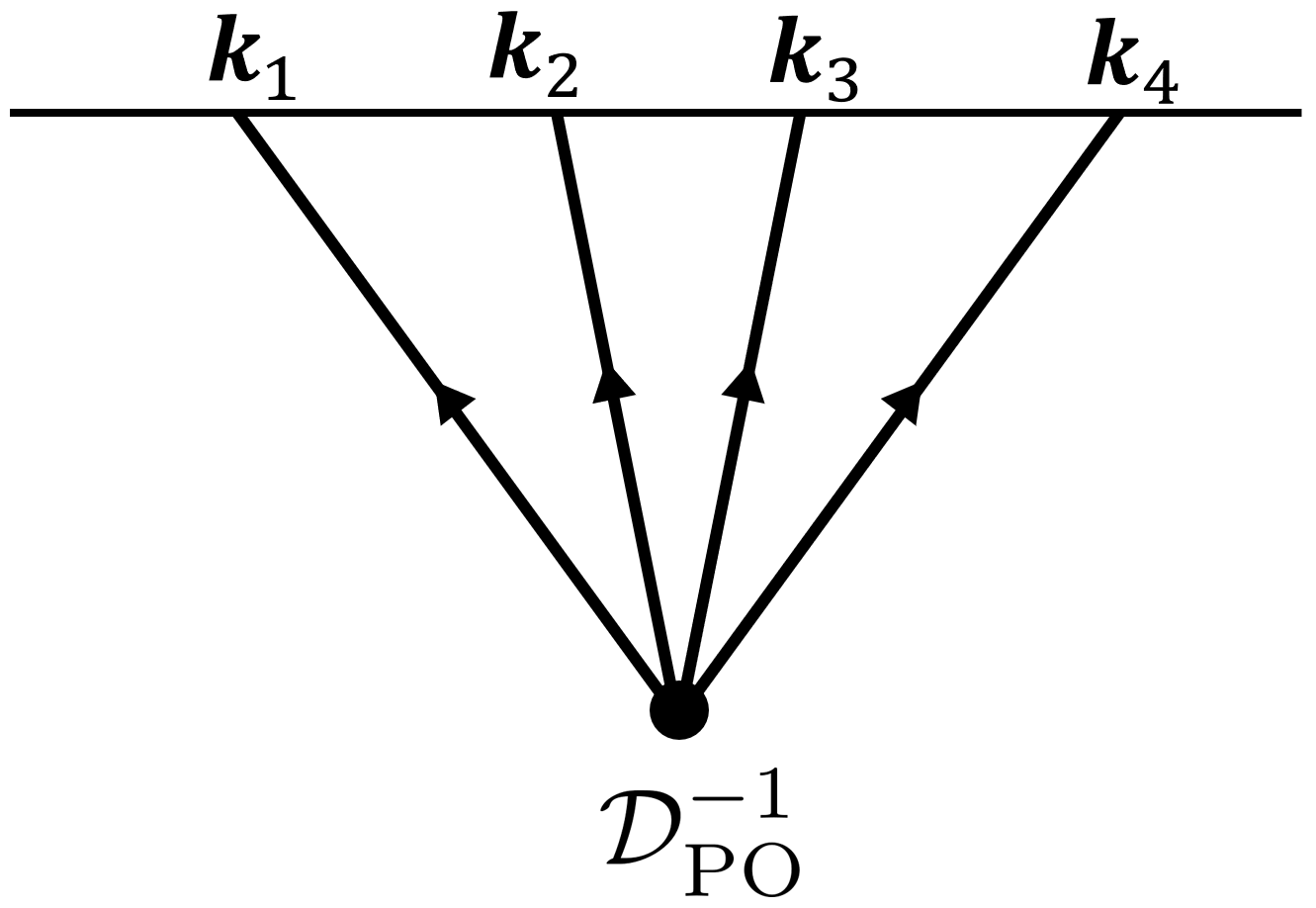}\\
	\caption{The contact diagram from the non-local interaction generating a parity-odd trispectrum. The black dot labeled by $\mathcal{D}_{\text{PO}}^{-1}$ indicates that the interaction is local in time but non-local in space.}
	\label{fig: ContactDiagram}
\end{figure}

\begin{figure}[htbp]
	\centering
	\includegraphics[width=0.3\textwidth]{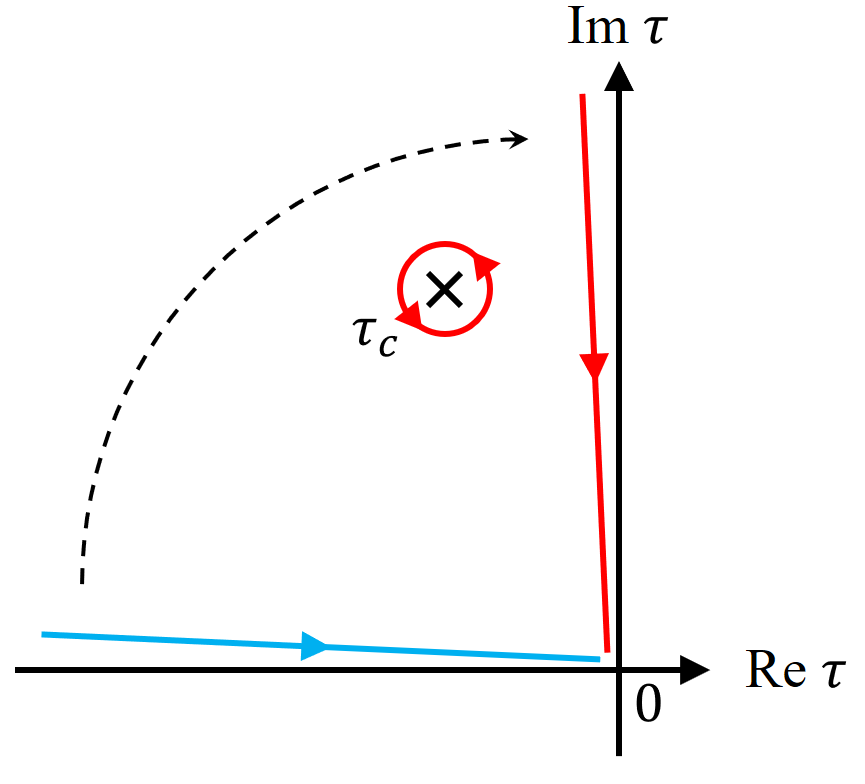}
	\caption{Representation of the $\tau$-complex plane and the Wick rotation in the non-local EFT. The blue contour is the original integration path specified by the $i\epsilon$-prescription. After Wick-rotating it to the red contour, one needs to pick up the pole at $\tau_c$. The contour integral along the imaginary axis is purely real and drops out of the result. The only non-zero contribution to the parity-odd trispectrum emerges from the residue at the pole $\tau=\tau_c$.}
	\label{fig: WickIllustration}
\end{figure}

With the non-local single-field EFT (\ref{eq: IRLag}), one can readily work out the resulting trispectrum, whose parity-odd part will be of our main interest. Let us switch to conformal coordinates with $a(\tau) = -1/(H\tau)$. For the $s$-channel contribution, the exchanged momentum is $\bm{s} \equiv \bm{k}_1+\bm{k}_2$, and the 4-point correlator is simply computed as a tree-level contact diagram using the Schwinger-Keldysh formalism (see Fig.~\ref{fig: ContactDiagram}),
\begin{align}
	\label{eq: 4ptIntegrand}
	\nonumber\langle\pi_c^4\rangle_{s,\mathrm{PO}}'=&-\frac{1}{\Lambda^6}\, \bm{s}\cdot(\bm{k}_1\times\bm{k}_3)(\bm{k}_1\cdot\bm{k}_2)(\bm{k}_3\cdot\bm{k}_4)\\
	\nonumber&\times 2i\,\Im \Bigg\{\int_{-\infty}^{0}d\tau \left(a^{-1}\mathcal{K}^+_1\overleftrightarrow{\partial_\tau}\mathcal{K}^+_2\right)\\
	\nonumber&\qquad\quad\Bigg[\frac{2\kappa a}{(s^2+m^2a^2)^2-4\kappa^2a^2s^2}+\mathcal{O}(\partial_\tau^2) \Bigg]\\
	&\qquad\quad\left(a^{-1}\mathcal{K}^+_3\overleftrightarrow{\partial_\tau}\mathcal{K}^+_4\right)\Bigg\}\,,
\end{align}
where $f\overleftrightarrow{\partial_\tau}g\equiv f\partial_{\tau}g-g\partial_{\tau}f$. $\mathcal{K}^+_{\sf{a}}\equiv u_{k_{\sf{a}}}(0)u_{k_{\sf{a}}}^*(\tau)$ is the bulk-to-boundary propagator of the Goldstone with the mode function
\begin{align}
	\label{eq: pi mode function}
	u_{k_{\sf{a}}}(\tau)=\frac{H}{\sqrt{2c_s^3 k_{\sf{a}}^3}}(1+ic_s k_{\sf{a}}\tau)e^{-i c_s k_{\sf{a}}\tau}~,
\end{align}
with ${\sf a}=1,\cdots,4$. A prime on a correlator indicates that we have dropped the momentum conserving delta function $(2\pi)^3\delta^3(\bm{k}_1 + \cdots + \bm{k}_4)$. Notice that the integrand in Eq.~(\ref{eq: 4ptIntegrand}) is analytic in the entire $\tau$-complex plane except at the root of the denominator
\begin{align}
	[s^2+m^2a^2(\tau_c)]^2-4\kappa^2a^2(\tau_c)s^2 = 0 \,.
\end{align}
This singularity in the upper left half complex plane lies at
\begin{align}
	\label{eq: PoleLocation}
	\tau_c \equiv \frac{1}{s}\left(-\frac{\kappa}{H} + \frac{im}{H}\sqrt{1 - \frac{\kappa^2}{m^2}}\right)\,.
\end{align}
In the limit $\kappa/H\rightarrow 0$, this pole is purely imaginary. As a result, the effect of the chemical potential is to shift the location of this pole away from the imaginary axis, as shown in Fig.~\ref{fig: WickIllustration}.
At LO, $\tau_c$ is a simple pole of the integrand while it is a double pole for the NLO term. In App.~\ref{sec: HOSingularitiesApp}, we also show that all higher-order terms in the non-local EFT (\ref{eq: IRLag}) share the same pole location at $\tau_c$, except that these poles are of increasing orders. As mentioned before, this essential singularity at $\tau_c$ is the manifestation of emergent non-locality in the IR theory.

\vskip 4pt
Implementing the $i\epsilon$-prescription, we now perform a Wick rotation on the upper half $\tau$-complex plane, $\tau=-e^{-i(\frac{\pi}{2} - \epsilon)}x$ with $x>0$, while taking into account the residue at $\tau_c$. This procedure is illustrated in Fig.~\ref{fig: WickIllustration}. The arc at infinity vanishes due to the exponential decay of the integrand. It is easy to check that the various ingredients transform as follows under the Wick rotation
\begin{align}
	\nonumber\int_{-\infty}^0\d\tau&\to -i\int_{0}^\infty \d x \in i\, \mathbb{R}\,,\\
	\nonumber \partial_\tau&\to -i\partial_x \in i\,\mathbbm{R}\,,\\
	\nonumber a(\tau)&\to a(ix)=\frac{i}{H x}\in i\,\mathbb{R}\,,\\
	\mathcal{K}_{\sf{a}}(\tau)&\to\mathcal{K}_{\sf{a}}(ix)=\frac{H^2}{2c_s^3 k_{\sf{a}}^3}(1+c_s k_{\sf{a}}x)e^{-c_s k_{\sf{a}}x}\in \mathbb{R}\,.
 \raisetag{45pt}
\end{align}
Interestingly, the Wick rotated integral is purely real and therefore vanishes from $\langle\pi_c^4\rangle'_{s,\mathrm{PO}}$. The only remaining contribution comes from the \textit{residue} at the singularity $\tau_c$, which reads
\begin{align}	
	\label{eq: 4ptResidue}
	\nonumber\langle\pi_c^4\rangle'_{s,\mathrm{PO}}=&-\frac{1}{\Lambda^6}\, \bm{s}\cdot(\bm{k}_1\times\bm{k}_3)(\bm{k}_1\cdot\bm{k}_2)(\bm{k}_3\cdot\bm{k}_4)\\
	\nonumber& \times 2i\,\Im\Bigg\{2\pi i \, \Res_{\tau\to \tau_c}\left(a^{-1}\mathcal{K}^+_1\overleftrightarrow{\partial_\tau}\mathcal{K}^+_2\right)\\
	\nonumber&\qquad\quad\left[\frac{2\kappa a}{(s^2+m^2a^2)^2-4\kappa^2a^2s^2}+\mathcal{O}(\partial_\tau^2)\right]\\
	&\qquad\quad\left(a^{-1}\mathcal{K}^+_3\overleftrightarrow{\partial_\tau}\mathcal{K}^+_4\right)\Bigg\}\,.
\end{align}
The vanishing of the Wick rotated integral can be understood as a consequence of the no-go theorem of parity violation in single-field theories. After Wick rotation but before actually computing the integral, one can perform a further spatial gradient expansion in powers of $s/(ma),s/(\kappa a)$, which corresponds to a \textit{local} single-field EFT. The no-go theorem forbids the presence of parity violation in such local single-field EFTs, thus the spatial gradient expansion must vanish term-by-term upon performing the integral and taking the imaginary part. Our result bypasses the no-go theorem precisely via \textit{emergent non-locality} at the singularity, which is derived from the fact that the UV theory is multi-field in nature. 

\vskip 4pt
To proceed to the final trispectrum, we relate the Goldstone boson to the curvature perturbation by using $\zeta = -H c_s^{3/2}f_\pi^{-2}\pi_c$, and introduce the dimensionless trispectrum $\calT$ by
\begin{align}
	\langle\zeta_{\bm{k}_1}\zeta_{\bm{k}_2}\zeta_{\bm{k}_3}\zeta_{\bm{k}_4}\rangle'=(2\pi)^6\Delta_\zeta^6\frac{(k_T/4)^3}{(k_1 k_2 k_3 k_4)^3}\calT(\bm{k}_1,\bm{k}_2,\bm{k}_3,\bm{k}_4)\,,
\end{align}
where $k_T\equiv \sum_{\sf a} k_{\sf a}$ is referred to as the total energy. The parity-odd dimensionless trispectrum at LO can be written as
\begin{align}
	\label{eq: PO-trispectrum}
	\nonumber \mathcal{T}_{\text{PO}}(\bm{k}_1,\bm{k}_2,\bm{k}_3,\bm{k}_4)&=\mathcal{A}_{\text{PO}}\, \Pi_s(\{\bm{k}_{\sf a}\})\, F_s(\{k_{\sf a}\},s)\\
	&\quad+\text{($t,u$ channels)}~,
\end{align}
where the amplitude $\mathcal{A}_{\text{PO}}$, the polarization factor $\Pi_s$ and the function $F_s$ read
\begin{align}
	\nonumber \mathcal{A}_{\text{PO}}&\equiv \frac{i}{(2\pi)^2 c_s^2 \Delta_\zeta^2}\left(\frac{H}{\Lambda}\right)^6\left(\frac{\kappa}{H}\right)\left(\frac{m}{2H}\right)^4~,\\
	\nonumber\Pi_s(\{\bm{k}_{\sf a}\})&\equiv \bm{s}\cdot(\bm{k}_1\times\bm{k}_3)\,(\bm{k}_1\cdot\bm{k}_2)\,(\bm{k}_3\cdot\bm{k}_4)~,\\
	\nonumber F_s(\{k_{\sf a}\},s)&\equiv\frac{64(k_1-k_2)(k_3-k_4)}{k_T^3s^2}\,\Im\Bigg[e^{i c_s k_T \tau_c}\\
	&\frac{i\pi (H\tau_c)^{6}(ik_{12}+c_s k_1 k_2 \tau_c)(ik_{34}+c_s k_3 k_4 \tau_c)}{m^4 \kappa\left(\kappa+sH\tau_c\right)}\Bigg]\,,
 \raisetag{80pt}
\end{align}
where $k_{12}\equiv k_1+k_2$, $k_{34}\equiv k_3+k_4$ and $\tau_c$ is the location of the singularity given in Eq.~(\ref{eq: PoleLocation}). The function $F_s$, when combined with $\Pi_s$, has been defined to be of order unity in regular kinematic configurations when $c_s \kappa, c_s m\ll H$. As such, it can be identified as a shape function for the parity-odd trispectrum. Note though that one cannot simply treat the prefactor $\mathcal{A}_{\text{PO}}$ as the amplitude of the parity-odd trispectrum, due to the dependence of $F_s(\{k_{\sf a}\},s)$ on the parameters $m,\kappa$. The overall size of the signal should rather be determined by looking at the full dimensionless trispectrum $\mathcal{T}_{\text{PO}}$.

\vskip 4pt
We now make a few comments on the analytical structure of the non-local EFT result. First, we see that the prefactor $\mathcal{A}_{\text{PO}}$ is imaginary\footnote{When splitting the trispectrum signal into parity-even and parity-odd contributions $\mathcal{T} = \mathcal{T}_{\text{PE}} + \mathcal{T}_{\text{PO}}$, the parity-odd contribution is identified as the imaginary part of the trispectrum $\mathcal{T}_{\text{PO}} = i \, \text{Im} \mathcal{T}$.} and the kinematic factor $\Pi_s$ enforces the parity-odd trispectrum to vanish in non-planar configurations, as for any $n$-point correlators with parity violation. Second, the function $F_s(\{k_{\sf a}\},s)$ encoding the inflationary dynamics is a combination of simple rational functions and exponential factors, which originates from products of Goldstone mode functions in the vertex integrand. This outcome is of course attributed to the residue theorem. Importantly, there is no total energy $k_T$-pole in the correlator $\langle\zeta^4\rangle'\propto k_T^3 F_s(\{k_{\sf a}\},s)$, suggesting that the parity-odd correlator can be written as a sum of factorized terms on the singularity at $\tau_c$.

\subsection{Shape Characteristics}

\begin{figure*}[ht]
	\centering
	\hspace*{-1cm}
	\includegraphics[width=1.08\textwidth]{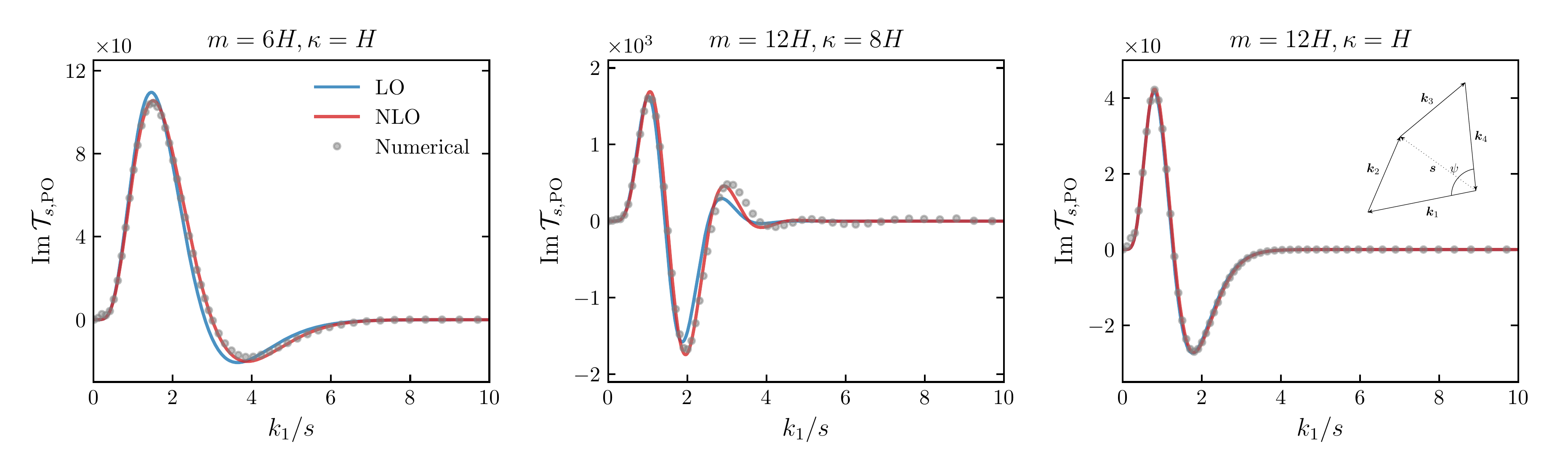}\\
	\caption{The $s$-channel parity-odd trispectrum shape $\Im\mathcal{T}_{s, \text{PO}}$ as defined in Eq.~(\ref{eq: PO-trispectrum}) for the $\partial_j \dot{\pi}_c \partial_i \partial_j \pi_c \sigma_i$ interaction in the non-planar tetrahedron kinematics configuration, with $k_1=k_3$, $k_2=k_4=\sqrt{s^2+k_1^2}$ and $\psi = \pi/3$ being the dihedral angle from the $(\bm k_1,\bm k_2)$-plane to the $(\bm k_3,\bm k_4)$-plane. The parameters are chosen as $m/H=6, \kappa/H=1$ \textit{(left)}, $m/H=12, \kappa/H=8$ \textit{(middle)}, and $m/H=12, \kappa/H=1$ \textit{(right)}, together with a common sound speed $c_s=0.1$ and $\Lambda=30H$. The \textcolor{pyblue}{blue} and \textcolor{pyred}{red} curves show the leading-order (LO) and the next-to-leading-order (NLO) non-local EFT results, respectively, and the \textcolor{gray}{gray} dotted line shows the exact numerical result in the two-field UV theory. The first visible resonance peak near the regular kinematics configuration is the low-speed collider signature, see Eq.~(\ref{eq: low-speed collider resonance}). The second peak (valley) and the subsequent ones in the exponentially attenuated tail are attributed to the new class of oscillatory signals, periodic in $k_1/s$, see Eq.~(\ref{eq: pole exponential}). For $t$ and $u$ channels, the exchanged momenta are not soft,  the low-speed collider resonance is absent, and the corresponding signals are suppressed.}
	\label{fig: Trispectrum}
\end{figure*}

The parity-odd trispectrum in the non-local EFT exhibits several interesting features. They are based on the observation that the dynamical function $F_s(\{k_{\sf a}\},s)$ contains an exponential factor $e^{i c_s k_T \tau_c}$ with both real and imaginary parts,
\begin{align}
	\label{eq: pole exponential}
	e^{i c_s k_T \tau_c}=\exp\left(-\frac{k_T}{s} \sqrt{\alpha_m^2 - \alpha_\kappa^2} \right)\exp\left( \frac{-i\alpha_\kappa k_T}{s}\right)\,.
\end{align}
This exponential factor is inherited from the Goldstone dynamical phase at the singularity, and leads to both a low-speed resonance peak and a new family of oscillatory signals:

\begin{enumerate}
	\item[$\bullet$] First, the real part of the exponent dictates that the overall amplitude of the parity-odd trispectrum is $\mathcal{O}(e^{-c_s m/H})$, which is non-perturbative in $m^{-1}$ (recall that $\alpha_\kappa<\alpha_m= c_s m/H$). Consistent with the no-go theorem, if we were to locally integrate out the massive spinning field, it would lead to a vanishing parity-odd contribution. Alternatively, a non-vanishing parity-odd signal arising from the exchange of a massive field must be non-perturbative in the mass. Now that the sound speed is reduced, the signal strength becomes considerably larger than the half-Boltzmann suppression $\mathcal{O}(e^{-\pi m/H})$ in the conventional case with a unit sound speed. Note also that the shape function is exponentially attenuated in the internal ultra-soft limit $s\to 0$, creating a resonance in the mildly-soft kinematic configuration. More explicitly, a resonance peak emerges in the trispectrum as a result of the interplay between a polynomial rising head and a exponentially attenuating tail:
	\begin{align}
		|\mathcal{T}_{s,\text{PO}}| \sim \left(\frac{k_S}{k_L}\right)^n e^{-\sqrt{\alpha_m^2-\alpha_\kappa^2}\,\frac{4k_S}{k_L}}\,,
	\end{align}
	where $n\simeq 2$ is a power-law proxy for the polynomial dependence, while $k_L\sim s$ is the long mode and $k_S\sim k_1,\cdots,k_4$ represents the short modes. Demanding $\partial |\mathcal{T}_{s,\text{PO}}|/\partial(k_S/k_L)=0$ yields the location of the resonance peak, given by
	\begin{align}
		\label{eq: low-speed collider resonance}
		\frac{k_L}{k_S}\sim\frac{4}{n}\sqrt{\alpha_m^2-\alpha_\kappa^2}\,.
	\end{align}
	This is essentially the same resonance peak as in the original low-speed collider framework, which lies at $k_L/k_S\sim \alpha_m = c_s m/H$ in the absence of chemical potential \cite{Jazayeri:2022kjy,Jazayeri:2023xcj}.
	
	\item[$\bullet$] On the other hand, the imaginary part in the exponential (\ref{eq: pole exponential}) gives a new class of oscillatory signals. The oscillation is periodic when expressed in the momentum ratio $k_T/s$, rather than logarithmic as in conventional cosmological collider signals. The oscillation frequency is determined not by the mass, but by the Goldstone sound speed and chemical potential through the dimensionless parameter $\alpha_\kappa = c_s \kappa/H$. Such oscillations are accompanied by the exponential decay in the limit $k_T/s\to \infty$, as noted above. It is interesting that even after integrating out the massive vector field, oscillatory features still appear within the IR non-local EFT. In contrast, conventional cosmological collider signals are \textit{not} captured by the non-local EFT result, as they encode the super-horizon \textit{on-shell} dynamics of the heavy field. One can think of the difference between the two types of oscillatory signals as a consequence of different dynamical phases in the complex $\tau$-plane. We will elaborate more on this understanding using the saddle-point approximation in a future work.
\end{enumerate}

\begin{figure}[htbp]
	\centering
	\includegraphics[width=0.5\textwidth]{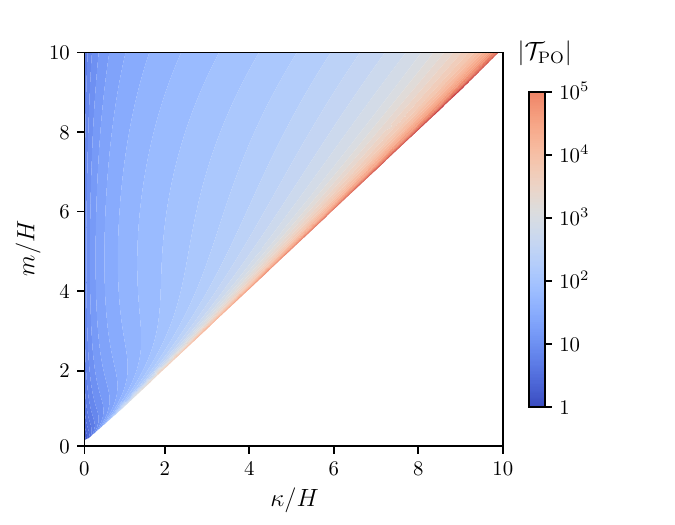}
	\caption{The size of the parity-odd trispectrum $|\calT_{\text{PO}}|$ generated by the interaction $\partial_j \dot{\pi}_c \partial_i \partial_j \pi_c \sigma_i$. The contours give the peak value of $\mathcal{T}_{\text{PO}}$ in logarithmic scale. The horizontal axis is the chemical potential $\kappa/H$ while the vertical axis shows the mass of the vector field $m/H$, both in Hubble units. The sound speed is chosen to be $c_s=0.1$, along with the cutoff scale $\Lambda/H=30$. The region $\kappa>m$ is excluded as the massive spinning particle develops a tachyonic instability, see Sec.~\ref{sec: Chemical Potential in the EFT of Inflation}.}
	\label{fig: SignalReach}
\end{figure}

\vskip 4pt
To confirm the two features above, we present the $s$-channel dimensionless parity-odd trispectrum $\text{Im}\, \calT_{s, \text{PO}}$ in Fig.~\ref{fig: Trispectrum} along with the numerical exact result computed with the UV theory (see App.~\ref{sec: ExactUVApp} for more details). As one can clearly see from the figure, there are multiple peaks and valleys among which the first one is the low-speed collider resonance and the others come from the new family of oscillatory signals. For a small sound speed $c_s \ll 1$ and chemical potential $\kappa/H \ll 1$, the LO non-local EFT result perfectly agrees with the exact one. Deviations start to appear when the chemical potential becomes comparable to the mass $\kappa \lesssim m$. In this case, the conventional cosmological collider signal becomes larger, dominating the collapsed limit (see the oscillating tail present in the exact result that is absent in the non-local EFT in the middle panel of Fig.~\ref{fig: Trispectrum}). The non-local EFT result nonetheless predicts well the other features of the signal.
We also show the signal strength of the parity-odd trispectrum in Fig.~\ref{fig: SignalReach}, where we find that our model can give sizable signals within the reach of future LSS observations. 

\section{Conclusion}
\label{sec: Conclusion}

Parity-violating cosmological correlators serve as a compelling evidence for new physics in the early universe. In this paper, we have shown that a chemical-potential driven dynamics of an additional massive spin-1 field during inflation generates a large parity-odd trispectrum in the presence of a reduced speed of sound for the Goldstone boson of broken time translations. The reduced sound speed limit admits a single-field non-local EFT description, in which the massive vector field is integrated out, leaving parity-violating self-interactions of the Goldstone boson of broken time translations that are local in time but non-local in space. Remarkably, despite the single-field content of the non-local EFT, this emergent spatial non-locality leads to a non-vanishing parity-odd trispectrum. We have shown that this correlator is completely determined as the residue at the singularity of the non-local interactions, bypassing the single-field no-go theorem on parity violation. Inspecting the characteristics of the parity-odd trispectrum, we have identified a low-speed-collider resonance peak in mildly-soft kinematics configurations and a new type of oscillatory patterns toward the collapsed limit that is distinct from the traditional cosmological collider signals. The exploration of the parameter space reveals a potentially large parity-odd signal, making it a noteworthy consideration for future observations.

\vskip 4pt
While in this work we have focused exclusively on parity violation in the non-local EFT, there are many interesting physics to explore in the UV theory, providing ample opportunities for future research. To name a few, the reduced Goldstone sound speed opens up new phase-space regions ($u\equiv s/(c_s k_{12})>1$, $v\equiv s/(c_s k_{34})>1$) for the trispectrum with new families of cosmological collider signals. It would be interesting to classify these signals and understand how they connect to the new oscillatory patterns lying in the internal soft-limit of the trispectrum found in the present work. Another important theoretical question is to give an exact solution to the low-speed collider trispectrum in the presence of chemical potential, especially in the new kinematics region $u,v>1$ where all the known solutions cease to apply. Since the chemical potential and the speed of sound are the only possible modifications of the dispersion relation up to quadratic order in a derivative expansion, the full exact solution would be marked as the last piece of the puzzle of solving boost-breaking tree-level trispectra. We hope to address these questions in a forthcoming work.

\begin{acknowledgments}
	
	We wish to thank Paolo Creminelli, David Stefanyszyn, Yi Wang and Masahide Yamaguchi for helpful discussions. XT would like to thank the Institut d'Astrophysique de Paris for kind hospitality. SJ, SRP and DW are supported by the European Research Council under the European Union's Horizon 2020 research and innovation programme (grant agreement No 758792, Starting Grant project GEODESI). XT and YZ are supported in part by
the National Key R$\&$D Program of China (No. 2021YFC2203100). This article is distributed under the Creative Commons Attribution International Licence (\href{https://creativecommons.org/licenses/by/4.0/}{CC-BY 4.0}).
	
\end{acknowledgments}

\appendix

\section{Singularity at Higher Orders}
\label{sec: HOSingularitiesApp}

At LO, the location $\tau_c$ \eqref{eq: PoleLocation} in the $\tau$-complex plane appears to be the only singularity of the integrand (\ref{eq: 4ptIntegrand}), where it manifests itself as a simple pole. In this appendix, we show that $\tau_c$ is the only singularity at \textit{any} finite order in the parity-odd sector of the series expansion (\ref{eq: timeDerivativeExpansion}), so that the parity-odd trispectrum is, to all orders, fixed by the singularity at $\tau_c$.

\vskip 4pt
In Fourier space, consider a further decomposition of $\D^{-1}_{\mathrm{PE}}$ in Eq.~(\ref{eq: non-local PE/PO operators}) into an identity part and a longitudinal part,
\begin{align}
	\D^{-1}_{\mathrm{PE}}=\D^{-1}_{\mathrm{I}}+\D^{-1}_{\mathrm{L}}\,,
\end{align}
with
\begin{align}
	[\D^{-1}_{\mathrm{I}}]_{ij}&\equiv\frac{s^2/a^2+m^2}{(s^2/a^2+m^2)^2-4\kappa^2 s^2/a^2}\,\delta_{ij}\,,\\
	[\D^{-1}_{\mathrm{L}}]_{ij}&\equiv\frac{-4\kappa^2}{(s^2/a^2+m^2)^2-4\kappa^2 s^2/a^2}\,\frac{s_i s_j/a^2}{s^2/a^2+m^2}\,.
\end{align}
Now both $\D^{-1}_{\mathrm{I}}$ and $\D^{-1}_{\mathrm{PO}}$ possess a pole at $\tau_c$, while the longitudinal part $\D^{-1}_{\mathrm{L}}$ has an extra pole at
\begin{align}
	\tau_L\equiv \frac{im}{sH}\,.
\end{align}
Thus we need to show that in the full expansion\footnote{Notice that we have applied the identity $a^{-2}\partial_\tau^2=\partial_t^2+H\partial_t$ to Eq.~(\ref{eq: timeDerivativeExpansion}).}
\begin{align}
	\nonumber&\left(\frac{\partial_\tau^2}{a^2}+\D\right)^{-1}=\,\\
	&\left(\D^{-1}_{\mathrm{I}}+\D^{-1}_{\mathrm{L}}+\D^{-1}_{\mathrm{PO}}\right)\sum_{n=0}^\infty\left\{-\frac{\partial_\tau^2}{a^2}\left(\D^{-1}_{\mathrm{I}}+\D^{-1}_{\mathrm{L}}+\D^{-1}_{\mathrm{PO}}\right)\right\}^{n}\,,
 \raisetag{55pt}
\end{align}
this extra pole never appears in the parity-odd sector. In fact, this is simply guaranteed by the tensorial structure. Namely, the parity-odd sector must have at least one factor of $\D^{-1}_{\mathrm{PO}}$:
\begin{align}
	\cdots \frac{\partial_\tau^2}{a^2}\, \D^{-1}_{\mathrm{PO}}\, \frac{\partial_\tau^2}{a^2}\cdots\,.
\end{align}
Now one can always find the \textit{closest} insertion of $[\D^{-1}]_{\mathrm{L}}$, with a sequence of $[\D^{-1}]_{\mathrm{I}}$ inserted in between,
\begin{align}
	\cdots \D^{-1}_{\mathrm{PO}}\, \frac{\partial_\tau^2}{a^2}\, \D^{-1}_{\mathrm{I}}\cdots\D^{-1}_{\mathrm{I}}\, \frac{\partial_\tau^2}{a^2}\, \D^{-1}_{\mathrm{L}}\, \cdots\,.
\end{align}
The tensor structure then reads
\begin{align}
	\cdots 2i\kappa \epsilon_{ijl} s_l \,\delta_{jm_{1}}\cdots \delta_{m_{2q}n} \, s_{n}s_{k}\cdots = 0\,.
\end{align}
Therefore, any insertion of $\D^{-1}_{\mathrm{L}}$ with its extra pole is eliminated by the nearest factor $\D^{-1}_{\mathrm{PO}}$. Henceforth, there is no extra $\tau_L$-singularity in the parity-odd sector of the non-local EFT. 

\vskip 4pt
Since finite steps of summation, multiplication and taking derivatives also do not generate new singularities, but only serve to increase the order of the $\tau_c$-pole, we conclude that $\tau_c$ is the only singularity of the non-local EFT at any finite order, and that the resulting parity-odd trispectrum only receives contribution from the $\tau_c$-pole.

\section{Details on the UV Computation}
\label{sec: ExactUVApp}

\begin{figure}[htbp!]
	\centering
	\includegraphics[width=0.3\textwidth]{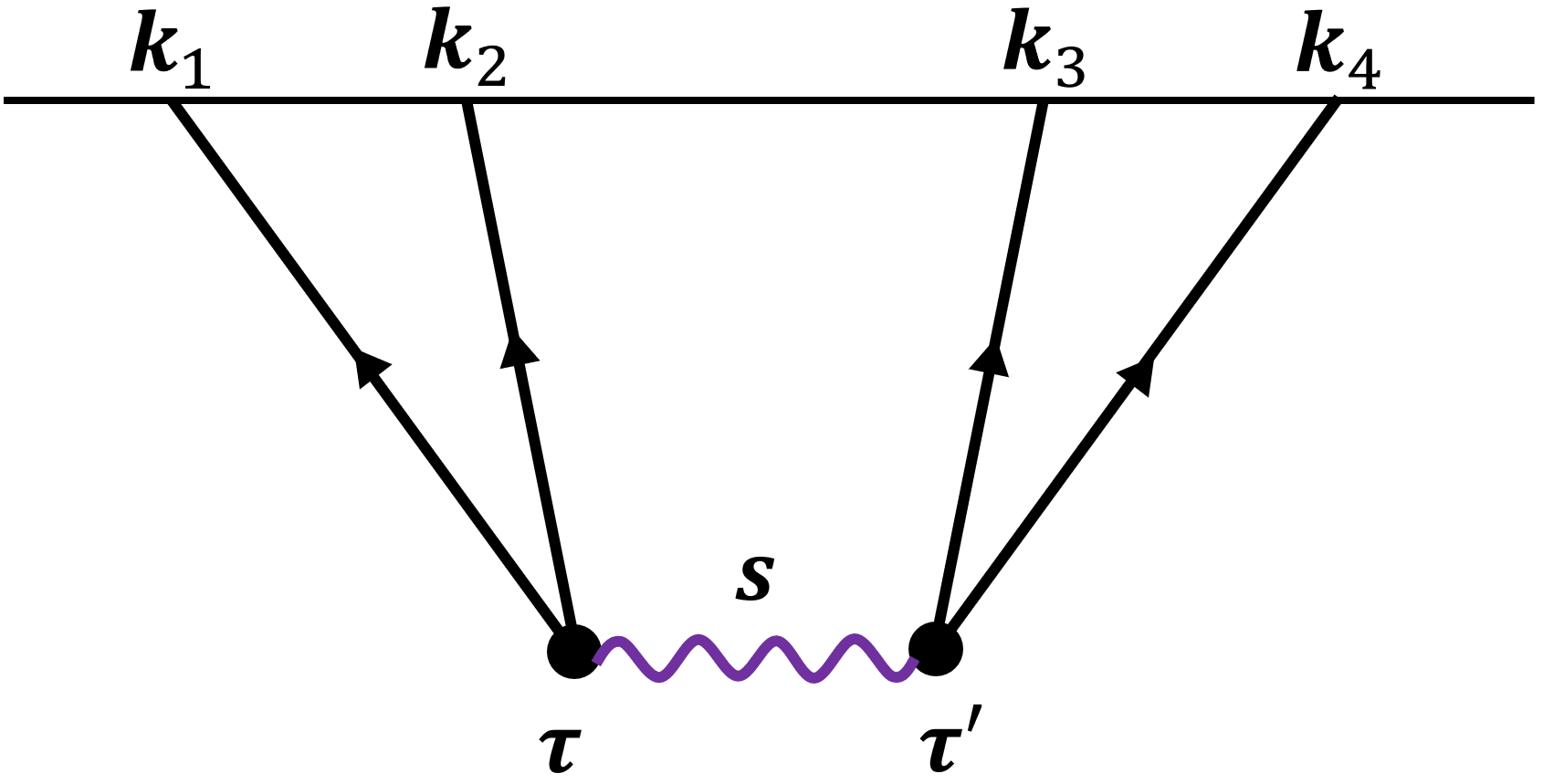}\\
	\caption{The trispectrum diagram in the UV theory. In this case both vertices are local in spacetime but they are connected by a propagating vector field, represented by the purple wavy line here.}
	\label{fig: ExchangeDiagram}
\end{figure}

In this appendix, we provide details on the exact calculation of the parity-odd trispectrum in the UV theory. Using the Schwinger-Keldysh diagrammatics, the $s$-channel diagram, see Fig.~\ref{fig: ExchangeDiagram}, from transverse vector exchange reads
\begin{align}
	\nonumber&\langle \pi_c^4\rangle_{s,\mathrm{T}}'=\frac{1}{\Lambda^6}(\bm{k}_1\cdot\bm{k}_2)(\bm{k}_3\cdot\bm{k}_4)\sum_{\lambda,\rm a,\rm b=\pm}(\bm{k}_1\cdot\bm{\varepsilon}^{*\lambda})(\bm{k}_3\cdot\bm{\varepsilon}^\lambda)\\
	&{\rm a\rm b}\int d\tau d\tau' \left(a^{-1}\mathcal{K}_1^{\rm a}\overleftrightarrow{\partial_\tau}\mathcal{K}_2^{\rm a}\right)\mathcal{G}^{\rm a\rm b}_\lambda(s,\tau,\tau')\left(a^{\prime-1}\mathcal{K}_3^{\rm b}\overleftrightarrow{\partial_{\tau'}}\mathcal{K}_4^{\rm b}\right)\,,
\end{align}
where we have abbreviated $a'\equiv a(\tau')$. Here, the polarization vector is 
\begin{align}
	\bm{\varepsilon}^\lambda(\hat{\bm{s}})=\frac{\hat{\bm{n}}-(\hat{\bm{n}}\cdot\hat{\bm{s}})\hat{\bm{s}}+i\lambda \,\hat{\bm{s}}\times\hat{\bm{n}}}{\sqrt{2[1-(\hat{\bm{n}}\cdot\hat{\bm{s}})^2]}}\,,
\end{align}
where $\hat{\bm{n}}$ is an arbitrary unit vector not parallel to $\hat{\bm{s}}$. The Goldstone mode function, associated with the bulk-to-boundary propagator $\mathcal{K}^{+}_{\sf{a}}$, has been defined in (\ref{eq: pi mode function}), its anti-time-ordered cousin reads
\begin{align}
	\mathcal{K}^{-}_{\sf{a}}&= \left(\mathcal{K}^{+}_{\sf{a}}\right)^*\,.
\end{align}
The vector propagators are 
\begin{align}
	\nonumber\mathcal{G}^{++}_\lambda(s,\tau,\tau')&=\theta(\tau-\tau')\sigma_{\lambda, k}(\tau)\sigma^*_{\lambda, k}(\tau')\\
	&\,+\theta(\tau'-\tau)\sigma^*_{\lambda, k}(\tau)\sigma_{\lambda, k}(\tau')\,,\\
	\mathcal{G}^{-+}_\lambda(s,\tau,\tau')&=\sigma_{\lambda, k}(\tau)\sigma^*_{\lambda, k}(\tau')~,\\
	\mathcal{G}^{+-}_\lambda(s,\tau,\tau')&=\left[\mathcal{G}^{-+}_\lambda(s,\tau,\tau')\right]^*~,\\
	\mathcal{G}^{--}_\lambda(s,\tau,\tau')&=\left[\mathcal{G}^{++}_\lambda(s,\tau,\tau')\right]^*\,,
\end{align}
where the mode function is expressed in terms of the Whittaker $W$-function,
\begin{align}
	\nonumber&\sigma_{\lambda, k}(\tau)=\frac{e^{-\lambda\pi\tilde{\kappa}/2}}{\sqrt{2k}}W_{i\lambda\tilde{\kappa},i\mu}(2ik\tau)~,\\
	&\text{with}\quad\mu\equiv\sqrt{\frac{m^2}{H^2}-\frac{1}{4}}~,\quad \tilde{\kappa}\equiv\frac{\kappa}{H}~.
\end{align}
The nested time integrals are solved numerically. We will provide analytical results for this correlator in a future work. The exact numerical result from the UV theory is contrasted with the non-local EFT one in Fig.~\ref{fig: Trispectrum}.

\bibliography{Refs}

\begin{thebibliography}{56}%
\makeatletter
\providecommand \@ifxundefined [1]{%
 \@ifx{#1\undefined}
}%
\providecommand \@ifnum [1]{%
 \ifnum #1\expandafter \@firstoftwo
 \else \expandafter \@secondoftwo
 \fi
}%
\providecommand \@ifx [1]{%
 \ifx #1\expandafter \@firstoftwo
 \else \expandafter \@secondoftwo
 \fi
}%
\providecommand \natexlab [1]{#1}%
\providecommand \enquote  [1]{``#1''}%
\providecommand \bibnamefont  [1]{#1}%
\providecommand \bibfnamefont [1]{#1}%
\providecommand \citenamefont [1]{#1}%
\providecommand \href@noop [0]{\@secondoftwo}%
\providecommand \href [0]{\begingroup \@sanitize@url \@href}%
\providecommand \@href[1]{\@@startlink{#1}\@@href}%
\providecommand \@@href[1]{\endgroup#1\@@endlink}%
\providecommand \@sanitize@url [0]{\catcode `\\12\catcode `\$12\catcode
  `\&12\catcode `\#12\catcode `\^12\catcode `\_12\catcode `\%12\relax}%
\providecommand \@@startlink[1]{}%
\providecommand \@@endlink[0]{}%
\providecommand \url  [0]{\begingroup\@sanitize@url \@url }%
\providecommand \@url [1]{\endgroup\@href {#1}{\urlprefix }}%
\providecommand \urlprefix  [0]{URL }%
\providecommand \Eprint [0]{\href }%
\providecommand \doibase [0]{https://doi.org/}%
\providecommand \selectlanguage [0]{\@gobble}%
\providecommand \bibinfo  [0]{\@secondoftwo}%
\providecommand \bibfield  [0]{\@secondoftwo}%
\providecommand \translation [1]{[#1]}%
\providecommand \BibitemOpen [0]{}%
\providecommand \bibitemStop [0]{}%
\providecommand \bibitemNoStop [0]{.\EOS\space}%
\providecommand \EOS [0]{\spacefactor3000\relax}%
\providecommand \BibitemShut  [1]{\csname bibitem#1\endcsname}%
\let\auto@bib@innerbib\@empty
\bibitem [{\citenamefont {Lee}\ and\ \citenamefont {Yang}(1956)}]{Lee:1956qn}%
  \BibitemOpen
  \bibfield  {author} {\bibinfo {author} {\bibfnamefont {T.~D.}\ \bibnamefont
  {Lee}}\ and\ \bibinfo {author} {\bibfnamefont {C.-N.}\ \bibnamefont {Yang}},\
  }\bibfield  {title} {\bibinfo {title} {{Question of Parity Conservation in
  Weak Interactions}},\ }\href {https://doi.org/10.1103/PhysRev.104.254}
  {\bibfield  {journal} {\bibinfo  {journal} {Phys. Rev.}\ }\textbf {\bibinfo
  {volume} {104}},\ \bibinfo {pages} {254} (\bibinfo {year}
  {1956})}\BibitemShut {NoStop}%
\bibitem [{\citenamefont {Wu}\ \emph {et~al.}(1957)\citenamefont {Wu},
  \citenamefont {Ambler}, \citenamefont {Hayward}, \citenamefont {Hoppes},\
  and\ \citenamefont {Hudson}}]{Wu:1957my}%
  \BibitemOpen
  \bibfield  {author} {\bibinfo {author} {\bibfnamefont {C.~S.}\ \bibnamefont
  {Wu}}, \bibinfo {author} {\bibfnamefont {E.}~\bibnamefont {Ambler}}, \bibinfo
  {author} {\bibfnamefont {R.~W.}\ \bibnamefont {Hayward}}, \bibinfo {author}
  {\bibfnamefont {D.~D.}\ \bibnamefont {Hoppes}},\ and\ \bibinfo {author}
  {\bibfnamefont {R.~P.}\ \bibnamefont {Hudson}},\ }\bibfield  {title}
  {\bibinfo {title} {{Experimental Test of Parity Conservation in $\beta$
  Decay}},\ }\href {https://doi.org/10.1103/PhysRev.105.1413} {\bibfield
  {journal} {\bibinfo  {journal} {Phys. Rev.}\ }\textbf {\bibinfo {volume}
  {105}},\ \bibinfo {pages} {1413} (\bibinfo {year} {1957})}\BibitemShut
  {NoStop}%
\bibitem [{\citenamefont {Glashow}(1961)}]{Glashow:1961tr}%
  \BibitemOpen
  \bibfield  {author} {\bibinfo {author} {\bibfnamefont {S.~L.}\ \bibnamefont
  {Glashow}},\ }\bibfield  {title} {\bibinfo {title} {{Partial Symmetries of
  Weak Interactions}},\ }\href {https://doi.org/10.1016/0029-5582(61)90469-2}
  {\bibfield  {journal} {\bibinfo  {journal} {Nucl. Phys.}\ }\textbf {\bibinfo
  {volume} {22}},\ \bibinfo {pages} {579} (\bibinfo {year} {1961})}\BibitemShut
  {NoStop}%
\bibitem [{\citenamefont {Weinberg}(1967)}]{Weinberg:1967tq}%
  \BibitemOpen
  \bibfield  {author} {\bibinfo {author} {\bibfnamefont {S.}~\bibnamefont
  {Weinberg}},\ }\bibfield  {title} {\bibinfo {title} {{A Model of Leptons}},\
  }\href {https://doi.org/10.1103/PhysRevLett.19.1264} {\bibfield  {journal}
  {\bibinfo  {journal} {Phys. Rev. Lett.}\ }\textbf {\bibinfo {volume} {19}},\
  \bibinfo {pages} {1264} (\bibinfo {year} {1967})}\BibitemShut {NoStop}%
\bibitem [{\citenamefont {Salam}(1968)}]{Salam:1968rm}%
  \BibitemOpen
  \bibfield  {author} {\bibinfo {author} {\bibfnamefont {A.}~\bibnamefont
  {Salam}},\ }\bibfield  {title} {\bibinfo {title} {{Weak and Electromagnetic
  Interactions}},\ }\href {https://doi.org/10.1142/9789812795915_0034}
  {\bibfield  {journal} {\bibinfo  {journal} {Conf. Proc. C}\ }\textbf
  {\bibinfo {volume} {680519}},\ \bibinfo {pages} {367} (\bibinfo {year}
  {1968})}\BibitemShut {NoStop}%
\bibitem [{\citenamefont {Minami}\ and\ \citenamefont
  {Komatsu}(2020)}]{Minami:2020odp}%
  \BibitemOpen
  \bibfield  {author} {\bibinfo {author} {\bibfnamefont {Y.}~\bibnamefont
  {Minami}}\ and\ \bibinfo {author} {\bibfnamefont {E.}~\bibnamefont
  {Komatsu}},\ }\bibfield  {title} {\bibinfo {title} {{New Extraction of the
  Cosmic Birefringence from the Planck 2018 Polarization Data}},\ }\href
  {https://doi.org/10.1103/PhysRevLett.125.221301} {\bibfield  {journal}
  {\bibinfo  {journal} {Phys. Rev. Lett.}\ }\textbf {\bibinfo {volume} {125}},\
  \bibinfo {pages} {221301} (\bibinfo {year} {2020})},\ \Eprint
  {https://arxiv.org/abs/2011.11254} {arXiv:2011.11254 [astro-ph.CO]}
  \BibitemShut {NoStop}%
\bibitem [{\citenamefont {Diego-Palazuelos}\ \emph {et~al.}(2022)\citenamefont
  {Diego-Palazuelos} \emph {et~al.}}]{Diego-Palazuelos:2022dsq}%
  \BibitemOpen
  \bibfield  {author} {\bibinfo {author} {\bibfnamefont {P.}~\bibnamefont
  {Diego-Palazuelos}} \emph {et~al.},\ }\bibfield  {title} {\bibinfo {title}
  {{Cosmic Birefringence from the Planck Data Release 4}},\ }\href
  {https://doi.org/10.1103/PhysRevLett.128.091302} {\bibfield  {journal}
  {\bibinfo  {journal} {Phys. Rev. Lett.}\ }\textbf {\bibinfo {volume} {128}},\
  \bibinfo {pages} {091302} (\bibinfo {year} {2022})},\ \Eprint
  {https://arxiv.org/abs/2201.07682} {arXiv:2201.07682 [astro-ph.CO]}
  \BibitemShut {NoStop}%
\bibitem [{\citenamefont {Hou}\ \emph {et~al.}(2022)\citenamefont {Hou},
  \citenamefont {Slepian},\ and\ \citenamefont {Cahn}}]{Hou:2022wfj}%
  \BibitemOpen
  \bibfield  {author} {\bibinfo {author} {\bibfnamefont {J.}~\bibnamefont
  {Hou}}, \bibinfo {author} {\bibfnamefont {Z.}~\bibnamefont {Slepian}},\ and\
  \bibinfo {author} {\bibfnamefont {R.~N.}\ \bibnamefont {Cahn}},\ }\bibfield
  {title} {\bibinfo {title} {{Measurement of Parity-Odd Modes in the
  Large-Scale 4-Point Correlation Function of SDSS BOSS DR12 CMASS and LOWZ
  Galaxies}},\ }\href@noop {} {\  (\bibinfo {year} {2022})},\ \Eprint
  {https://arxiv.org/abs/2206.03625} {arXiv:2206.03625 [astro-ph.CO]}
  \BibitemShut {NoStop}%
\bibitem [{\citenamefont {Philcox}(2022)}]{Philcox:2022hkh}%
  \BibitemOpen
  \bibfield  {author} {\bibinfo {author} {\bibfnamefont {O.~H.~E.}\
  \bibnamefont {Philcox}},\ }\bibfield  {title} {\bibinfo {title} {{Probing
  parity violation with the four-point correlation function of BOSS
  galaxies}},\ }\href {https://doi.org/10.1103/PhysRevD.106.063501} {\bibfield
  {journal} {\bibinfo  {journal} {Phys. Rev. D}\ }\textbf {\bibinfo {volume}
  {106}},\ \bibinfo {pages} {063501} (\bibinfo {year} {2022})},\ \Eprint
  {https://arxiv.org/abs/2206.04227} {arXiv:2206.04227 [astro-ph.CO]}
  \BibitemShut {NoStop}%
\bibitem [{\citenamefont {Cabass}\ \emph
  {et~al.}(2023{\natexlab{a}})\citenamefont {Cabass}, \citenamefont {Ivanov},\
  and\ \citenamefont {Philcox}}]{Cabass:2022oap}%
  \BibitemOpen
  \bibfield  {author} {\bibinfo {author} {\bibfnamefont {G.}~\bibnamefont
  {Cabass}}, \bibinfo {author} {\bibfnamefont {M.~M.}\ \bibnamefont {Ivanov}},\
  and\ \bibinfo {author} {\bibfnamefont {O.~H.~E.}\ \bibnamefont {Philcox}},\
  }\bibfield  {title} {\bibinfo {title} {{Colliders and ghosts: Constraining
  inflation with the parity-odd galaxy four-point function}},\ }\href
  {https://doi.org/10.1103/PhysRevD.107.023523} {\bibfield  {journal} {\bibinfo
   {journal} {Phys. Rev. D}\ }\textbf {\bibinfo {volume} {107}},\ \bibinfo
  {pages} {023523} (\bibinfo {year} {2023}{\natexlab{a}})},\ \Eprint
  {https://arxiv.org/abs/2210.16320} {arXiv:2210.16320 [astro-ph.CO]}
  \BibitemShut {NoStop}%
\bibitem [{\citenamefont {Philcox}(2023)}]{Philcox:2023ffy}%
  \BibitemOpen
  \bibfield  {author} {\bibinfo {author} {\bibfnamefont {O.~H.~E.}\
  \bibnamefont {Philcox}},\ }\bibfield  {title} {\bibinfo {title} {{Do the CMB
  Temperature Fluctuations Conserve Parity?}},\ }\href@noop {} {\  (\bibinfo
  {year} {2023})},\ \Eprint {https://arxiv.org/abs/2303.12106}
  {arXiv:2303.12106 [astro-ph.CO]} \BibitemShut {NoStop}%
\bibitem [{\citenamefont {Philcox}\ and\ \citenamefont
  {Shiraishi}(2023)}]{Philcox:2023ypl}%
  \BibitemOpen
  \bibfield  {author} {\bibinfo {author} {\bibfnamefont {O.~H.~E.}\
  \bibnamefont {Philcox}}\ and\ \bibinfo {author} {\bibfnamefont
  {M.}~\bibnamefont {Shiraishi}},\ }\bibfield  {title} {\bibinfo {title}
  {{Testing Parity Symmetry with the Polarized Cosmic Microwave Background}},\
  }\href@noop {} {\  (\bibinfo {year} {2023})},\ \Eprint
  {https://arxiv.org/abs/2308.03831} {arXiv:2308.03831 [astro-ph.CO]}
  \BibitemShut {NoStop}%
\bibitem [{\citenamefont {Liu}\ \emph {et~al.}(2020)\citenamefont {Liu},
  \citenamefont {Tong}, \citenamefont {Wang},\ and\ \citenamefont
  {Xianyu}}]{Liu:2019fag}%
  \BibitemOpen
  \bibfield  {author} {\bibinfo {author} {\bibfnamefont {T.}~\bibnamefont
  {Liu}}, \bibinfo {author} {\bibfnamefont {X.}~\bibnamefont {Tong}}, \bibinfo
  {author} {\bibfnamefont {Y.}~\bibnamefont {Wang}},\ and\ \bibinfo {author}
  {\bibfnamefont {Z.-Z.}\ \bibnamefont {Xianyu}},\ }\bibfield  {title}
  {\bibinfo {title} {{Probing P and CP Violations on the Cosmological
  Collider}},\ }\href {https://doi.org/10.1007/JHEP04(2020)189} {\bibfield
  {journal} {\bibinfo  {journal} {JHEP}\ }\textbf {\bibinfo {volume} {04}},\
  \bibinfo {pages} {189}},\ \Eprint {https://arxiv.org/abs/1909.01819}
  {arXiv:1909.01819 [hep-ph]} \BibitemShut {NoStop}%
\bibitem [{\citenamefont {Cabass}\ \emph
  {et~al.}(2023{\natexlab{b}})\citenamefont {Cabass}, \citenamefont {Jazayeri},
  \citenamefont {Pajer},\ and\ \citenamefont {Stefanyszyn}}]{Cabass:2022rhr}%
  \BibitemOpen
  \bibfield  {author} {\bibinfo {author} {\bibfnamefont {G.}~\bibnamefont
  {Cabass}}, \bibinfo {author} {\bibfnamefont {S.}~\bibnamefont {Jazayeri}},
  \bibinfo {author} {\bibfnamefont {E.}~\bibnamefont {Pajer}},\ and\ \bibinfo
  {author} {\bibfnamefont {D.}~\bibnamefont {Stefanyszyn}},\ }\bibfield
  {title} {\bibinfo {title} {{Parity violation in the scalar trispectrum: no-go
  theorems and yes-go examples}},\ }\href
  {https://doi.org/10.1007/JHEP02(2023)021} {\bibfield  {journal} {\bibinfo
  {journal} {JHEP}\ }\textbf {\bibinfo {volume} {02}},\ \bibinfo {pages}
  {021}},\ \Eprint {https://arxiv.org/abs/2210.02907} {arXiv:2210.02907
  [hep-th]} \BibitemShut {NoStop}%
\bibitem [{\citenamefont {Creque-Sarbinowski}\ \emph
  {et~al.}(2023)\citenamefont {Creque-Sarbinowski}, \citenamefont {Alexander},
  \citenamefont {Kamionkowski},\ and\ \citenamefont
  {Philcox}}]{Creque-Sarbinowski:2023wmb}%
  \BibitemOpen
  \bibfield  {author} {\bibinfo {author} {\bibfnamefont {C.}~\bibnamefont
  {Creque-Sarbinowski}}, \bibinfo {author} {\bibfnamefont {S.}~\bibnamefont
  {Alexander}}, \bibinfo {author} {\bibfnamefont {M.}~\bibnamefont
  {Kamionkowski}},\ and\ \bibinfo {author} {\bibfnamefont {O.}~\bibnamefont
  {Philcox}},\ }\bibfield  {title} {\bibinfo {title} {{Parity-Violating
  Trispectrum from Chern-Simons Gravity}},\ }\href@noop {} {\  (\bibinfo {year}
  {2023})},\ \Eprint {https://arxiv.org/abs/2303.04815} {arXiv:2303.04815
  [astro-ph.CO]} \BibitemShut {NoStop}%
\bibitem [{\citenamefont {Lee}\ \emph {et~al.}(2023)\citenamefont {Lee},
  \citenamefont {McCulloch},\ and\ \citenamefont {Pajer}}]{Lee:2023jby}%
  \BibitemOpen
  \bibfield  {author} {\bibinfo {author} {\bibfnamefont {M.~H.~G.}\
  \bibnamefont {Lee}}, \bibinfo {author} {\bibfnamefont {C.}~\bibnamefont
  {McCulloch}},\ and\ \bibinfo {author} {\bibfnamefont {E.}~\bibnamefont
  {Pajer}},\ }\bibfield  {title} {\bibinfo {title} {{Leading Loops in
  Cosmological Correlators}},\ }\href@noop {} {\  (\bibinfo {year} {2023})},\
  \Eprint {https://arxiv.org/abs/2305.11228} {arXiv:2305.11228 [hep-th]}
  \BibitemShut {NoStop}%
\bibitem [{\citenamefont {Baumann}\ and\ \citenamefont
  {Green}(2012)}]{Baumann:2011nk}%
  \BibitemOpen
  \bibfield  {author} {\bibinfo {author} {\bibfnamefont {D.}~\bibnamefont
  {Baumann}}\ and\ \bibinfo {author} {\bibfnamefont {D.}~\bibnamefont
  {Green}},\ }\bibfield  {title} {\bibinfo {title} {{Signatures of
  Supersymmetry from the Early Universe}},\ }\href
  {https://doi.org/10.1103/PhysRevD.85.103520} {\bibfield  {journal} {\bibinfo
  {journal} {Phys. Rev. D}\ }\textbf {\bibinfo {volume} {85}},\ \bibinfo
  {pages} {103520} (\bibinfo {year} {2012})},\ \Eprint
  {https://arxiv.org/abs/1109.0292} {arXiv:1109.0292 [hep-th]} \BibitemShut
  {NoStop}%
\bibitem [{\citenamefont {Delacretaz}\ \emph {et~al.}(2017)\citenamefont
  {Delacretaz}, \citenamefont {Gorbenko},\ and\ \citenamefont
  {Senatore}}]{Delacretaz:2016nhw}%
  \BibitemOpen
  \bibfield  {author} {\bibinfo {author} {\bibfnamefont {L.~V.}\ \bibnamefont
  {Delacretaz}}, \bibinfo {author} {\bibfnamefont {V.}~\bibnamefont
  {Gorbenko}},\ and\ \bibinfo {author} {\bibfnamefont {L.}~\bibnamefont
  {Senatore}},\ }\bibfield  {title} {\bibinfo {title} {{The Supersymmetric
  Effective Field Theory of Inflation}},\ }\href
  {https://doi.org/10.1007/JHEP03(2017)063} {\bibfield  {journal} {\bibinfo
  {journal} {JHEP}\ }\textbf {\bibinfo {volume} {03}},\ \bibinfo {pages}
  {063}},\ \Eprint {https://arxiv.org/abs/1610.04227} {arXiv:1610.04227
  [hep-th]} \BibitemShut {NoStop}%
\bibitem [{\citenamefont {Alexander}\ \emph {et~al.}(2019)\citenamefont
  {Alexander}, \citenamefont {Gates}, \citenamefont {Jenks}, \citenamefont
  {Koutrolikos},\ and\ \citenamefont {McDonough}}]{Alexander:2019vtb}%
  \BibitemOpen
  \bibfield  {author} {\bibinfo {author} {\bibfnamefont {S.}~\bibnamefont
  {Alexander}}, \bibinfo {author} {\bibfnamefont {S.~J.}\ \bibnamefont
  {Gates}}, \bibinfo {author} {\bibfnamefont {L.}~\bibnamefont {Jenks}},
  \bibinfo {author} {\bibfnamefont {K.}~\bibnamefont {Koutrolikos}},\ and\
  \bibinfo {author} {\bibfnamefont {E.}~\bibnamefont {McDonough}},\ }\bibfield
  {title} {\bibinfo {title} {{Higher Spin Supersymmetry at the Cosmological
  Collider: Sculpting SUSY Rilles in the CMB}},\ }\href
  {https://doi.org/10.1007/JHEP10(2019)156} {\bibfield  {journal} {\bibinfo
  {journal} {JHEP}\ }\textbf {\bibinfo {volume} {10}},\ \bibinfo {pages}
  {156}},\ \Eprint {https://arxiv.org/abs/1907.05829} {arXiv:1907.05829
  [hep-th]} \BibitemShut {NoStop}%
\bibitem [{\citenamefont {Chen}\ \emph
  {et~al.}(2017{\natexlab{a}})\citenamefont {Chen}, \citenamefont {Wang},\ and\
  \citenamefont {Xianyu}}]{Chen:2016uwp}%
  \BibitemOpen
  \bibfield  {author} {\bibinfo {author} {\bibfnamefont {X.}~\bibnamefont
  {Chen}}, \bibinfo {author} {\bibfnamefont {Y.}~\bibnamefont {Wang}},\ and\
  \bibinfo {author} {\bibfnamefont {Z.-Z.}\ \bibnamefont {Xianyu}},\ }\bibfield
   {title} {\bibinfo {title} {{Standard Model Background of the Cosmological
  Collider}},\ }\href {https://doi.org/10.1103/PhysRevLett.118.261302}
  {\bibfield  {journal} {\bibinfo  {journal} {Phys. Rev. Lett.}\ }\textbf
  {\bibinfo {volume} {118}},\ \bibinfo {pages} {261302} (\bibinfo {year}
  {2017}{\natexlab{a}})},\ \Eprint {https://arxiv.org/abs/1610.06597}
  {arXiv:1610.06597 [hep-th]} \BibitemShut {NoStop}%
\bibitem [{\citenamefont {Kumar}\ and\ \citenamefont
  {Sundrum}(2018)}]{Kumar:2017ecc}%
  \BibitemOpen
  \bibfield  {author} {\bibinfo {author} {\bibfnamefont {S.}~\bibnamefont
  {Kumar}}\ and\ \bibinfo {author} {\bibfnamefont {R.}~\bibnamefont
  {Sundrum}},\ }\bibfield  {title} {\bibinfo {title} {{Heavy-Lifting of Gauge
  Theories By Cosmic Inflation}},\ }\href
  {https://doi.org/10.1007/JHEP05(2018)011} {\bibfield  {journal} {\bibinfo
  {journal} {JHEP}\ }\textbf {\bibinfo {volume} {05}},\ \bibinfo {pages}
  {011}},\ \Eprint {https://arxiv.org/abs/1711.03988} {arXiv:1711.03988
  [hep-ph]} \BibitemShut {NoStop}%
\bibitem [{\citenamefont {Hook}\ \emph {et~al.}(2020)\citenamefont {Hook},
  \citenamefont {Huang},\ and\ \citenamefont {Racco}}]{Hook:2019zxa}%
  \BibitemOpen
  \bibfield  {author} {\bibinfo {author} {\bibfnamefont {A.}~\bibnamefont
  {Hook}}, \bibinfo {author} {\bibfnamefont {J.}~\bibnamefont {Huang}},\ and\
  \bibinfo {author} {\bibfnamefont {D.}~\bibnamefont {Racco}},\ }\bibfield
  {title} {\bibinfo {title} {{Searches for other vacua. Part II. A new
  Higgstory at the cosmological collider}},\ }\href
  {https://doi.org/10.1007/JHEP01(2020)105} {\bibfield  {journal} {\bibinfo
  {journal} {JHEP}\ }\textbf {\bibinfo {volume} {01}},\ \bibinfo {pages}
  {105}},\ \Eprint {https://arxiv.org/abs/1907.10624} {arXiv:1907.10624
  [hep-ph]} \BibitemShut {NoStop}%
\bibitem [{\citenamefont {Chen}\ \emph {et~al.}(2016)\citenamefont {Chen},
  \citenamefont {Wang},\ and\ \citenamefont {Xianyu}}]{Chen:2016nrs}%
  \BibitemOpen
  \bibfield  {author} {\bibinfo {author} {\bibfnamefont {X.}~\bibnamefont
  {Chen}}, \bibinfo {author} {\bibfnamefont {Y.}~\bibnamefont {Wang}},\ and\
  \bibinfo {author} {\bibfnamefont {Z.-Z.}\ \bibnamefont {Xianyu}},\ }\bibfield
   {title} {\bibinfo {title} {{Loop Corrections to Standard Model Fields in
  Inflation}},\ }\href {https://doi.org/10.1007/JHEP08(2016)051} {\bibfield
  {journal} {\bibinfo  {journal} {JHEP}\ }\textbf {\bibinfo {volume} {08}},\
  \bibinfo {pages} {051}},\ \Eprint {https://arxiv.org/abs/1604.07841}
  {arXiv:1604.07841 [hep-th]} \BibitemShut {NoStop}%
\bibitem [{\citenamefont {Chen}\ \emph
  {et~al.}(2017{\natexlab{b}})\citenamefont {Chen}, \citenamefont {Wang},\ and\
  \citenamefont {Xianyu}}]{Chen:2016hrz}%
  \BibitemOpen
  \bibfield  {author} {\bibinfo {author} {\bibfnamefont {X.}~\bibnamefont
  {Chen}}, \bibinfo {author} {\bibfnamefont {Y.}~\bibnamefont {Wang}},\ and\
  \bibinfo {author} {\bibfnamefont {Z.-Z.}\ \bibnamefont {Xianyu}},\ }\bibfield
   {title} {\bibinfo {title} {{Standard Model Mass Spectrum in Inflationary
  Universe}},\ }\href {https://doi.org/10.1007/JHEP04(2017)058} {\bibfield
  {journal} {\bibinfo  {journal} {JHEP}\ }\textbf {\bibinfo {volume} {04}},\
  \bibinfo {pages} {058}},\ \Eprint {https://arxiv.org/abs/1612.08122}
  {arXiv:1612.08122 [hep-th]} \BibitemShut {NoStop}%
\bibitem [{\citenamefont {Chen}\ and\ \citenamefont
  {Wang}(2010)}]{Chen:2009zp}%
  \BibitemOpen
  \bibfield  {author} {\bibinfo {author} {\bibfnamefont {X.}~\bibnamefont
  {Chen}}\ and\ \bibinfo {author} {\bibfnamefont {Y.}~\bibnamefont {Wang}},\
  }\bibfield  {title} {\bibinfo {title} {{Quasi-Single Field Inflation and
  Non-Gaussianities}},\ }\href {https://doi.org/10.1088/1475-7516/2010/04/027}
  {\bibfield  {journal} {\bibinfo  {journal} {JCAP}\ }\textbf {\bibinfo
  {volume} {04}},\ \bibinfo {pages} {027}},\ \Eprint
  {https://arxiv.org/abs/0911.3380} {arXiv:0911.3380 [hep-th]} \BibitemShut
  {NoStop}%
\bibitem [{\citenamefont {McAllister}\ \emph {et~al.}(2012)\citenamefont
  {McAllister}, \citenamefont {Renaux-Petel},\ and\ \citenamefont
  {Xu}}]{McAllister:2012am}%
  \BibitemOpen
  \bibfield  {author} {\bibinfo {author} {\bibfnamefont {L.}~\bibnamefont
  {McAllister}}, \bibinfo {author} {\bibfnamefont {S.}~\bibnamefont
  {Renaux-Petel}},\ and\ \bibinfo {author} {\bibfnamefont {G.}~\bibnamefont
  {Xu}},\ }\bibfield  {title} {\bibinfo {title} {{A Statistical Approach to
  Multifield Inflation: Many-field Perturbations Beyond Slow Roll}},\ }\href
  {https://doi.org/10.1088/1475-7516/2012/10/046} {\bibfield  {journal}
  {\bibinfo  {journal} {JCAP}\ }\textbf {\bibinfo {volume} {10}},\ \bibinfo
  {pages} {046}},\ \Eprint {https://arxiv.org/abs/1207.0317} {arXiv:1207.0317
  [astro-ph.CO]} \BibitemShut {NoStop}%
\bibitem [{\citenamefont {Lu}(2022)}]{Lu:2021gso}%
  \BibitemOpen
  \bibfield  {author} {\bibinfo {author} {\bibfnamefont {S.}~\bibnamefont
  {Lu}},\ }\bibfield  {title} {\bibinfo {title} {{Axion isocurvature
  collider}},\ }\href {https://doi.org/10.1007/JHEP04(2022)157} {\bibfield
  {journal} {\bibinfo  {journal} {JHEP}\ }\textbf {\bibinfo {volume} {04}},\
  \bibinfo {pages} {157}},\ \Eprint {https://arxiv.org/abs/2103.05958}
  {arXiv:2103.05958 [hep-th]} \BibitemShut {NoStop}%
\bibitem [{\citenamefont {Chen}\ \emph {et~al.}(2023)\citenamefont {Chen},
  \citenamefont {Fan},\ and\ \citenamefont {Li}}]{Chen:2023txq}%
  \BibitemOpen
  \bibfield  {author} {\bibinfo {author} {\bibfnamefont {X.}~\bibnamefont
  {Chen}}, \bibinfo {author} {\bibfnamefont {J.}~\bibnamefont {Fan}},\ and\
  \bibinfo {author} {\bibfnamefont {L.}~\bibnamefont {Li}},\ }\bibfield
  {title} {\bibinfo {title} {{New inflationary probes of axion dark matter}},\
  }\href@noop {} {\  (\bibinfo {year} {2023})},\ \Eprint
  {https://arxiv.org/abs/2303.03406} {arXiv:2303.03406 [hep-ph]} \BibitemShut
  {NoStop}%
\bibitem [{\citenamefont {Rindani}\ and\ \citenamefont
  {Sivakumar}(1985)}]{Rindani:1985pi}%
  \BibitemOpen
  \bibfield  {author} {\bibinfo {author} {\bibfnamefont {S.~D.}\ \bibnamefont
  {Rindani}}\ and\ \bibinfo {author} {\bibfnamefont {M.}~\bibnamefont
  {Sivakumar}},\ }\bibfield  {title} {\bibinfo {title} {{Gauge - Invariant
  Description of Massive Higher - Spin Particles by Dimensional Reduction}},\
  }\href {https://doi.org/10.1103/PhysRevD.32.3238} {\bibfield  {journal}
  {\bibinfo  {journal} {Phys. Rev. D}\ }\textbf {\bibinfo {volume} {32}},\
  \bibinfo {pages} {3238} (\bibinfo {year} {1985})}\BibitemShut {NoStop}%
\bibitem [{\citenamefont {Aragone}\ \emph {et~al.}(1987)\citenamefont
  {Aragone}, \citenamefont {Deser},\ and\ \citenamefont
  {Yang}}]{Aragone:1987dtt}%
  \BibitemOpen
  \bibfield  {author} {\bibinfo {author} {\bibfnamefont {C.}~\bibnamefont
  {Aragone}}, \bibinfo {author} {\bibfnamefont {S.}~\bibnamefont {Deser}},\
  and\ \bibinfo {author} {\bibfnamefont {Z.}~\bibnamefont {Yang}},\ }\bibfield
  {title} {\bibinfo {title} {{Massive Higher Spin From Dimensional Reduction of
  Gauge Fields}},\ }\href {https://doi.org/10.1016/S0003-4916(87)80005-2}
  {\bibfield  {journal} {\bibinfo  {journal} {Annals Phys.}\ }\textbf {\bibinfo
  {volume} {179}},\ \bibinfo {pages} {76} (\bibinfo {year} {1987})}\BibitemShut
  {NoStop}%
\bibitem [{\citenamefont {Kumar}\ and\ \citenamefont
  {Sundrum}(2019)}]{Kumar:2018jxz}%
  \BibitemOpen
  \bibfield  {author} {\bibinfo {author} {\bibfnamefont {S.}~\bibnamefont
  {Kumar}}\ and\ \bibinfo {author} {\bibfnamefont {R.}~\bibnamefont
  {Sundrum}},\ }\bibfield  {title} {\bibinfo {title} {{Seeing
  Higher-Dimensional Grand Unification In Primordial Non-Gaussianities}},\
  }\href {https://doi.org/10.1007/JHEP04(2019)120} {\bibfield  {journal}
  {\bibinfo  {journal} {JHEP}\ }\textbf {\bibinfo {volume} {04}},\ \bibinfo
  {pages} {120}},\ \Eprint {https://arxiv.org/abs/1811.11200} {arXiv:1811.11200
  [hep-ph]} \BibitemShut {NoStop}%
\bibitem [{\citenamefont {Baumann}\ and\ \citenamefont
  {McAllister}(2015)}]{Baumann:2014nda}%
  \BibitemOpen
  \bibfield  {author} {\bibinfo {author} {\bibfnamefont {D.}~\bibnamefont
  {Baumann}}\ and\ \bibinfo {author} {\bibfnamefont {L.}~\bibnamefont
  {McAllister}},\ }\href {https://doi.org/10.1017/CBO9781316105733} {\emph
  {\bibinfo {title} {{Inflation and String Theory}}}},\ Cambridge Monographs on
  Mathematical Physics\ (\bibinfo  {publisher} {Cambridge University Press},\
  \bibinfo {year} {2015})\ \Eprint {https://arxiv.org/abs/1404.2601}
  {arXiv:1404.2601 [hep-th]} \BibitemShut {NoStop}%
\bibitem [{\citenamefont {Adshead}\ and\ \citenamefont
  {Sfakianakis}(2015)}]{Adshead:2015kza}%
  \BibitemOpen
  \bibfield  {author} {\bibinfo {author} {\bibfnamefont {P.}~\bibnamefont
  {Adshead}}\ and\ \bibinfo {author} {\bibfnamefont {E.~I.}\ \bibnamefont
  {Sfakianakis}},\ }\bibfield  {title} {\bibinfo {title} {{Fermion production
  during and after axion inflation}},\ }\href
  {https://doi.org/10.1088/1475-7516/2015/11/021} {\bibfield  {journal}
  {\bibinfo  {journal} {JCAP}\ }\textbf {\bibinfo {volume} {11}},\ \bibinfo
  {pages} {021}},\ \Eprint {https://arxiv.org/abs/1508.00891} {arXiv:1508.00891
  [hep-ph]} \BibitemShut {NoStop}%
\bibitem [{\citenamefont {Wang}\ and\ \citenamefont
  {Xianyu}(2020{\natexlab{a}})}]{Wang:2019gbi}%
  \BibitemOpen
  \bibfield  {author} {\bibinfo {author} {\bibfnamefont {L.-T.}\ \bibnamefont
  {Wang}}\ and\ \bibinfo {author} {\bibfnamefont {Z.-Z.}\ \bibnamefont
  {Xianyu}},\ }\bibfield  {title} {\bibinfo {title} {{In Search of Large
  Signals at the Cosmological Collider}},\ }\href
  {https://doi.org/10.1007/JHEP02(2020)044} {\bibfield  {journal} {\bibinfo
  {journal} {JHEP}\ }\textbf {\bibinfo {volume} {02}},\ \bibinfo {pages}
  {044}},\ \Eprint {https://arxiv.org/abs/1910.12876} {arXiv:1910.12876
  [hep-ph]} \BibitemShut {NoStop}%
\bibitem [{\citenamefont {Wang}\ and\ \citenamefont
  {Xianyu}(2020{\natexlab{b}})}]{Wang:2020ioa}%
  \BibitemOpen
  \bibfield  {author} {\bibinfo {author} {\bibfnamefont {L.-T.}\ \bibnamefont
  {Wang}}\ and\ \bibinfo {author} {\bibfnamefont {Z.-Z.}\ \bibnamefont
  {Xianyu}},\ }\bibfield  {title} {\bibinfo {title} {{Gauge Boson Signals at
  the Cosmological Collider}},\ }\href
  {https://doi.org/10.1007/JHEP11(2020)082} {\bibfield  {journal} {\bibinfo
  {journal} {JHEP}\ }\textbf {\bibinfo {volume} {11}},\ \bibinfo {pages}
  {082}},\ \Eprint {https://arxiv.org/abs/2004.02887} {arXiv:2004.02887
  [hep-ph]} \BibitemShut {NoStop}%
\bibitem [{\citenamefont {Sou}\ \emph {et~al.}(2021)\citenamefont {Sou},
  \citenamefont {Tong},\ and\ \citenamefont {Wang}}]{Sou:2021juh}%
  \BibitemOpen
  \bibfield  {author} {\bibinfo {author} {\bibfnamefont {C.~M.}\ \bibnamefont
  {Sou}}, \bibinfo {author} {\bibfnamefont {X.}~\bibnamefont {Tong}},\ and\
  \bibinfo {author} {\bibfnamefont {Y.}~\bibnamefont {Wang}},\ }\bibfield
  {title} {\bibinfo {title} {{Chemical-potential-assisted particle production
  in FRW spacetimes}},\ }\href {https://doi.org/10.1007/JHEP06(2021)129}
  {\bibfield  {journal} {\bibinfo  {journal} {JHEP}\ }\textbf {\bibinfo
  {volume} {06}},\ \bibinfo {pages} {129}},\ \Eprint
  {https://arxiv.org/abs/2104.08772} {arXiv:2104.08772 [hep-th]} \BibitemShut
  {NoStop}%
\bibitem [{\citenamefont {Wang}\ \emph {et~al.}(2022)\citenamefont {Wang},
  \citenamefont {Xianyu},\ and\ \citenamefont {Zhong}}]{Wang:2021qez}%
  \BibitemOpen
  \bibfield  {author} {\bibinfo {author} {\bibfnamefont {L.-T.}\ \bibnamefont
  {Wang}}, \bibinfo {author} {\bibfnamefont {Z.-Z.}\ \bibnamefont {Xianyu}},\
  and\ \bibinfo {author} {\bibfnamefont {Y.-M.}\ \bibnamefont {Zhong}},\
  }\bibfield  {title} {\bibinfo {title} {{Precision calculation of inflation
  correlators at one loop}},\ }\href {https://doi.org/10.1007/JHEP02(2022)085}
  {\bibfield  {journal} {\bibinfo  {journal} {JHEP}\ }\textbf {\bibinfo
  {volume} {02}},\ \bibinfo {pages} {085}},\ \Eprint
  {https://arxiv.org/abs/2109.14635} {arXiv:2109.14635 [hep-ph]} \BibitemShut
  {NoStop}%
\bibitem [{\citenamefont {Tong}\ and\ \citenamefont
  {Xianyu}(2022)}]{Tong:2022cdz}%
  \BibitemOpen
  \bibfield  {author} {\bibinfo {author} {\bibfnamefont {X.}~\bibnamefont
  {Tong}}\ and\ \bibinfo {author} {\bibfnamefont {Z.-Z.}\ \bibnamefont
  {Xianyu}},\ }\bibfield  {title} {\bibinfo {title} {{Large spin-2 signals at
  the cosmological collider}},\ }\href
  {https://doi.org/10.1007/JHEP10(2022)194} {\bibfield  {journal} {\bibinfo
  {journal} {JHEP}\ }\textbf {\bibinfo {volume} {10}},\ \bibinfo {pages}
  {194}},\ \Eprint {https://arxiv.org/abs/2203.06349} {arXiv:2203.06349
  [hep-ph]} \BibitemShut {NoStop}%
\bibitem [{\citenamefont {Qin}\ and\ \citenamefont
  {Xianyu}(2022)}]{Qin:2022lva}%
  \BibitemOpen
  \bibfield  {author} {\bibinfo {author} {\bibfnamefont {Z.}~\bibnamefont
  {Qin}}\ and\ \bibinfo {author} {\bibfnamefont {Z.-Z.}\ \bibnamefont
  {Xianyu}},\ }\bibfield  {title} {\bibinfo {title} {{Phase information in
  cosmological collider signals}},\ }\href
  {https://doi.org/10.1007/JHEP10(2022)192} {\bibfield  {journal} {\bibinfo
  {journal} {JHEP}\ }\textbf {\bibinfo {volume} {10}},\ \bibinfo {pages}
  {192}},\ \Eprint {https://arxiv.org/abs/2205.01692} {arXiv:2205.01692
  [hep-th]} \BibitemShut {NoStop}%
\bibitem [{\citenamefont {Qin}\ and\ \citenamefont
  {Xianyu}(2023)}]{Qin:2022fbv}%
  \BibitemOpen
  \bibfield  {author} {\bibinfo {author} {\bibfnamefont {Z.}~\bibnamefont
  {Qin}}\ and\ \bibinfo {author} {\bibfnamefont {Z.-Z.}\ \bibnamefont
  {Xianyu}},\ }\bibfield  {title} {\bibinfo {title} {{Helical inflation
  correlators: partial Mellin-Barnes and bootstrap equations}},\ }\href
  {https://doi.org/10.1007/JHEP04(2023)059} {\bibfield  {journal} {\bibinfo
  {journal} {JHEP}\ }\textbf {\bibinfo {volume} {04}},\ \bibinfo {pages}
  {059}},\ \Eprint {https://arxiv.org/abs/2208.13790} {arXiv:2208.13790
  [hep-th]} \BibitemShut {NoStop}%
\bibitem [{\citenamefont {Jazayeri}\ and\ \citenamefont
  {Renaux-Petel}(2022)}]{Jazayeri:2022kjy}%
  \BibitemOpen
  \bibfield  {author} {\bibinfo {author} {\bibfnamefont {S.}~\bibnamefont
  {Jazayeri}}\ and\ \bibinfo {author} {\bibfnamefont {S.}~\bibnamefont
  {Renaux-Petel}},\ }\bibfield  {title} {\bibinfo {title} {{Cosmological
  bootstrap in slow motion}},\ }\href {https://doi.org/10.1007/JHEP12(2022)137}
  {\bibfield  {journal} {\bibinfo  {journal} {JHEP}\ }\textbf {\bibinfo
  {volume} {12}},\ \bibinfo {pages} {137}},\ \Eprint
  {https://arxiv.org/abs/2205.10340} {arXiv:2205.10340 [hep-th]} \BibitemShut
  {NoStop}%
\bibitem [{\citenamefont {Jazayeri}\ \emph {et~al.}(2023)\citenamefont
  {Jazayeri}, \citenamefont {Renaux-Petel},\ and\ \citenamefont
  {Werth}}]{Jazayeri:2023xcj}%
  \BibitemOpen
  \bibfield  {author} {\bibinfo {author} {\bibfnamefont {S.}~\bibnamefont
  {Jazayeri}}, \bibinfo {author} {\bibfnamefont {S.}~\bibnamefont
  {Renaux-Petel}},\ and\ \bibinfo {author} {\bibfnamefont {D.}~\bibnamefont
  {Werth}},\ }\bibfield  {title} {\bibinfo {title} {{Shapes of the Cosmological
  Low-Speed Collider}},\ }\href@noop {} {\  (\bibinfo {year} {2023})},\ \Eprint
  {https://arxiv.org/abs/2307.01751} {arXiv:2307.01751 [hep-th]} \BibitemShut
  {NoStop}%
\bibitem [{\citenamefont {Cheung}\ \emph {et~al.}(2008)\citenamefont {Cheung},
  \citenamefont {Creminelli}, \citenamefont {Fitzpatrick}, \citenamefont
  {Kaplan},\ and\ \citenamefont {Senatore}}]{Cheung:2007st}%
  \BibitemOpen
  \bibfield  {author} {\bibinfo {author} {\bibfnamefont {C.}~\bibnamefont
  {Cheung}}, \bibinfo {author} {\bibfnamefont {P.}~\bibnamefont {Creminelli}},
  \bibinfo {author} {\bibfnamefont {A.~L.}\ \bibnamefont {Fitzpatrick}},
  \bibinfo {author} {\bibfnamefont {J.}~\bibnamefont {Kaplan}},\ and\ \bibinfo
  {author} {\bibfnamefont {L.}~\bibnamefont {Senatore}},\ }\bibfield  {title}
  {\bibinfo {title} {{The Effective Field Theory of Inflation}},\ }\href
  {https://doi.org/10.1088/1126-6708/2008/03/014} {\bibfield  {journal}
  {\bibinfo  {journal} {JHEP}\ }\textbf {\bibinfo {volume} {03}},\ \bibinfo
  {pages} {014}},\ \Eprint {https://arxiv.org/abs/0709.0293} {arXiv:0709.0293
  [hep-th]} \BibitemShut {NoStop}%
\bibitem [{\citenamefont {Senatore}\ and\ \citenamefont
  {Zaldarriaga}(2012)}]{Senatore:2010wk}%
  \BibitemOpen
  \bibfield  {author} {\bibinfo {author} {\bibfnamefont {L.}~\bibnamefont
  {Senatore}}\ and\ \bibinfo {author} {\bibfnamefont {M.}~\bibnamefont
  {Zaldarriaga}},\ }\bibfield  {title} {\bibinfo {title} {{The Effective Field
  Theory of Multifield Inflation}},\ }\href
  {https://doi.org/10.1007/JHEP04(2012)024} {\bibfield  {journal} {\bibinfo
  {journal} {JHEP}\ }\textbf {\bibinfo {volume} {04}},\ \bibinfo {pages}
  {024}},\ \Eprint {https://arxiv.org/abs/1009.2093} {arXiv:1009.2093 [hep-th]}
  \BibitemShut {NoStop}%
\bibitem [{\citenamefont {Piazza}\ and\ \citenamefont
  {Vernizzi}(2013)}]{Piazza:2013coa}%
  \BibitemOpen
  \bibfield  {author} {\bibinfo {author} {\bibfnamefont {F.}~\bibnamefont
  {Piazza}}\ and\ \bibinfo {author} {\bibfnamefont {F.}~\bibnamefont
  {Vernizzi}},\ }\bibfield  {title} {\bibinfo {title} {{Effective Field Theory
  of Cosmological Perturbations}},\ }\href
  {https://doi.org/10.1088/0264-9381/30/21/214007} {\bibfield  {journal}
  {\bibinfo  {journal} {Class. Quant. Grav.}\ }\textbf {\bibinfo {volume}
  {30}},\ \bibinfo {pages} {214007} (\bibinfo {year} {2013})},\ \Eprint
  {https://arxiv.org/abs/1307.4350} {arXiv:1307.4350 [hep-th]} \BibitemShut
  {NoStop}%
\bibitem [{\citenamefont {Anber}\ and\ \citenamefont
  {Sorbo}(2010)}]{Anber:2009ua}%
  \BibitemOpen
  \bibfield  {author} {\bibinfo {author} {\bibfnamefont {M.~M.}\ \bibnamefont
  {Anber}}\ and\ \bibinfo {author} {\bibfnamefont {L.}~\bibnamefont {Sorbo}},\
  }\bibfield  {title} {\bibinfo {title} {{Naturally inflating on steep
  potentials through electromagnetic dissipation}},\ }\href
  {https://doi.org/10.1103/PhysRevD.81.043534} {\bibfield  {journal} {\bibinfo
  {journal} {Phys. Rev. D}\ }\textbf {\bibinfo {volume} {81}},\ \bibinfo
  {pages} {043534} (\bibinfo {year} {2010})},\ \Eprint
  {https://arxiv.org/abs/0908.4089} {arXiv:0908.4089 [hep-th]} \BibitemShut
  {NoStop}%
\bibitem [{\citenamefont {Barnaby}\ and\ \citenamefont
  {Peloso}(2011)}]{Barnaby:2010vf}%
  \BibitemOpen
  \bibfield  {author} {\bibinfo {author} {\bibfnamefont {N.}~\bibnamefont
  {Barnaby}}\ and\ \bibinfo {author} {\bibfnamefont {M.}~\bibnamefont
  {Peloso}},\ }\bibfield  {title} {\bibinfo {title} {{Large Nongaussianity in
  Axion Inflation}},\ }\href {https://doi.org/10.1103/PhysRevLett.106.181301}
  {\bibfield  {journal} {\bibinfo  {journal} {Phys. Rev. Lett.}\ }\textbf
  {\bibinfo {volume} {106}},\ \bibinfo {pages} {181301} (\bibinfo {year}
  {2011})},\ \Eprint {https://arxiv.org/abs/1011.1500} {arXiv:1011.1500
  [hep-ph]} \BibitemShut {NoStop}%
\bibitem [{\citenamefont {Sorbo}(2011)}]{Sorbo:2011rz}%
  \BibitemOpen
  \bibfield  {author} {\bibinfo {author} {\bibfnamefont {L.}~\bibnamefont
  {Sorbo}},\ }\bibfield  {title} {\bibinfo {title} {{Parity violation in the
  Cosmic Microwave Background from a pseudoscalar inflaton}},\ }\href
  {https://doi.org/10.1088/1475-7516/2011/06/003} {\bibfield  {journal}
  {\bibinfo  {journal} {JCAP}\ }\textbf {\bibinfo {volume} {06}},\ \bibinfo
  {pages} {003}},\ \Eprint {https://arxiv.org/abs/1101.1525} {arXiv:1101.1525
  [astro-ph.CO]} \BibitemShut {NoStop}%
\bibitem [{\citenamefont {Barnaby}\ \emph {et~al.}(2012)\citenamefont
  {Barnaby}, \citenamefont {Pajer},\ and\ \citenamefont
  {Peloso}}]{Barnaby:2011qe}%
  \BibitemOpen
  \bibfield  {author} {\bibinfo {author} {\bibfnamefont {N.}~\bibnamefont
  {Barnaby}}, \bibinfo {author} {\bibfnamefont {E.}~\bibnamefont {Pajer}},\
  and\ \bibinfo {author} {\bibfnamefont {M.}~\bibnamefont {Peloso}},\
  }\bibfield  {title} {\bibinfo {title} {{Gauge Field Production in Axion
  Inflation: Consequences for Monodromy, non-Gaussianity in the CMB, and
  Gravitational Waves at Interferometers}},\ }\href
  {https://doi.org/10.1103/PhysRevD.85.023525} {\bibfield  {journal} {\bibinfo
  {journal} {Phys. Rev. D}\ }\textbf {\bibinfo {volume} {85}},\ \bibinfo
  {pages} {023525} (\bibinfo {year} {2012})},\ \Eprint
  {https://arxiv.org/abs/1110.3327} {arXiv:1110.3327 [astro-ph.CO]}
  \BibitemShut {NoStop}%
\bibitem [{\citenamefont {Pajer}\ and\ \citenamefont
  {Peloso}(2013)}]{Pajer:2013fsa}%
  \BibitemOpen
  \bibfield  {author} {\bibinfo {author} {\bibfnamefont {E.}~\bibnamefont
  {Pajer}}\ and\ \bibinfo {author} {\bibfnamefont {M.}~\bibnamefont {Peloso}},\
  }\bibfield  {title} {\bibinfo {title} {{A review of Axion Inflation in the
  era of Planck}},\ }\href {https://doi.org/10.1088/0264-9381/30/21/214002}
  {\bibfield  {journal} {\bibinfo  {journal} {Class. Quant. Grav.}\ }\textbf
  {\bibinfo {volume} {30}},\ \bibinfo {pages} {214002} (\bibinfo {year}
  {2013})},\ \Eprint {https://arxiv.org/abs/1305.3557} {arXiv:1305.3557
  [hep-th]} \BibitemShut {NoStop}%
\bibitem [{\citenamefont {Lee}\ \emph {et~al.}(2016)\citenamefont {Lee},
  \citenamefont {Baumann},\ and\ \citenamefont {Pimentel}}]{Lee:2016vti}%
  \BibitemOpen
  \bibfield  {author} {\bibinfo {author} {\bibfnamefont {H.}~\bibnamefont
  {Lee}}, \bibinfo {author} {\bibfnamefont {D.}~\bibnamefont {Baumann}},\ and\
  \bibinfo {author} {\bibfnamefont {G.~L.}\ \bibnamefont {Pimentel}},\
  }\bibfield  {title} {\bibinfo {title} {{Non-Gaussianity as a Particle
  Detector}},\ }\href {https://doi.org/10.1007/JHEP12(2016)040} {\bibfield
  {journal} {\bibinfo  {journal} {JHEP}\ }\textbf {\bibinfo {volume} {12}},\
  \bibinfo {pages} {040}},\ \Eprint {https://arxiv.org/abs/1607.03735}
  {arXiv:1607.03735 [hep-th]} \BibitemShut {NoStop}%
\bibitem [{\citenamefont {Baumann}\ and\ \citenamefont
  {Green}(2011)}]{Baumann:2011su}%
  \BibitemOpen
  \bibfield  {author} {\bibinfo {author} {\bibfnamefont {D.}~\bibnamefont
  {Baumann}}\ and\ \bibinfo {author} {\bibfnamefont {D.}~\bibnamefont
  {Green}},\ }\bibfield  {title} {\bibinfo {title} {{Equilateral
  Non-Gaussianity and New Physics on the Horizon}},\ }\href
  {https://doi.org/10.1088/1475-7516/2011/09/014} {\bibfield  {journal}
  {\bibinfo  {journal} {JCAP}\ }\textbf {\bibinfo {volume} {09}},\ \bibinfo
  {pages} {014}},\ \Eprint {https://arxiv.org/abs/1102.5343} {arXiv:1102.5343
  [hep-th]} \BibitemShut {NoStop}%
\bibitem [{\citenamefont {Baumann}\ \emph {et~al.}(2015)\citenamefont
  {Baumann}, \citenamefont {Green},\ and\ \citenamefont
  {Porto}}]{Baumann:2014cja}%
  \BibitemOpen
  \bibfield  {author} {\bibinfo {author} {\bibfnamefont {D.}~\bibnamefont
  {Baumann}}, \bibinfo {author} {\bibfnamefont {D.}~\bibnamefont {Green}},\
  and\ \bibinfo {author} {\bibfnamefont {R.~A.}\ \bibnamefont {Porto}},\
  }\bibfield  {title} {\bibinfo {title} {{B-modes and the Nature of
  Inflation}},\ }\href {https://doi.org/10.1088/1475-7516/2015/01/016}
  {\bibfield  {journal} {\bibinfo  {journal} {JCAP}\ }\textbf {\bibinfo
  {volume} {01}},\ \bibinfo {pages} {016}},\ \Eprint
  {https://arxiv.org/abs/1407.2621} {arXiv:1407.2621 [hep-th]} \BibitemShut
  {NoStop}%
\bibitem [{\citenamefont {Jazayeri}\ \emph {et~al.}(2021)\citenamefont
  {Jazayeri}, \citenamefont {Pajer},\ and\ \citenamefont
  {Stefanyszyn}}]{Jazayeri:2021fvk}%
  \BibitemOpen
  \bibfield  {author} {\bibinfo {author} {\bibfnamefont {S.}~\bibnamefont
  {Jazayeri}}, \bibinfo {author} {\bibfnamefont {E.}~\bibnamefont {Pajer}},\
  and\ \bibinfo {author} {\bibfnamefont {D.}~\bibnamefont {Stefanyszyn}},\
  }\bibfield  {title} {\bibinfo {title} {{From locality and unitarity to
  cosmological correlators}},\ }\href {https://doi.org/10.1007/JHEP10(2021)065}
  {\bibfield  {journal} {\bibinfo  {journal} {JHEP}\ }\textbf {\bibinfo
  {volume} {10}},\ \bibinfo {pages} {065}},\ \Eprint
  {https://arxiv.org/abs/2103.08649} {arXiv:2103.08649 [hep-th]} \BibitemShut
  {NoStop}%
\bibitem [{\citenamefont {Ade}\ \emph {et~al.}(2014)\citenamefont {Ade} \emph
  {et~al.}}]{Planck:2013wtn}%
  \BibitemOpen
  \bibfield  {author} {\bibinfo {author} {\bibfnamefont {P.~A.~R.}\
  \bibnamefont {Ade}} \emph {et~al.} (\bibinfo {collaboration} {Planck}),\
  }\bibfield  {title} {\bibinfo {title} {{Planck 2013 Results. XXIV.
  Constraints on primordial non-Gaussianity}},\ }\href
  {https://doi.org/10.1051/0004-6361/201321554} {\bibfield  {journal} {\bibinfo
   {journal} {Astron. Astrophys.}\ }\textbf {\bibinfo {volume} {571}},\
  \bibinfo {pages} {A24} (\bibinfo {year} {2014})},\ \Eprint
  {https://arxiv.org/abs/1303.5084} {arXiv:1303.5084 [astro-ph.CO]}
  \BibitemShut {NoStop}%
\bibitem [{\citenamefont {Akrami}\ \emph {et~al.}(2020)\citenamefont {Akrami}
  \emph {et~al.}}]{Planck:2019kim}%
  \BibitemOpen
  \bibfield  {author} {\bibinfo {author} {\bibfnamefont {Y.}~\bibnamefont
  {Akrami}} \emph {et~al.} (\bibinfo {collaboration} {Planck}),\ }\bibfield
  {title} {\bibinfo {title} {{Planck 2018 results. IX. Constraints on
  primordial non-Gaussianity}},\ }\href
  {https://doi.org/10.1051/0004-6361/201935891} {\bibfield  {journal} {\bibinfo
   {journal} {Astron. Astrophys.}\ }\textbf {\bibinfo {volume} {641}},\
  \bibinfo {pages} {A9} (\bibinfo {year} {2020})},\ \Eprint
  {https://arxiv.org/abs/1905.05697} {arXiv:1905.05697 [astro-ph.CO]}
  \BibitemShut {NoStop}%
\end{thebibliography}%

\end{document}